\documentclass[shortnames,nojss]{jss}

\usepackage[normalem]{ulem}
\usepackage{amsmath} 
\usepackage{amsthm} 
\usepackage{bm}
\usepackage{subcaption}
\usepackage{soul,xcolor}
\setstcolor{red}
\usepackage{rotating} 
\usepackage{multirow}
\usepackage[utf8]{inputenc}

\author{Thong Pham \\ RIKEN Center for AIP \And 
        Paul Sheridan \\ Hirosaki University \And
        Hidetoshi Shimodaira\\ Kyoto University \\
        RIKEN Center for AIP}
\title{\pkg{PAFit}: an \proglang{R} Package for the Non-Parametric Estimation of Preferential Attachment and Node Fitness in Temporal Complex Networks}

\Plainauthor{Thong Pham, Paul Sheridan, Hidetoshi Shimodaira} 
\Plaintitle{PAFit: an R Package for the Non-parametric Estimation of Preferential Attachment and Node Fitness in Temporal Complex Networks} 
\Shorttitle{\pkg{PAFit}: Non-Parametric Estimation of Preferential Attachment and Node Fitness} 

\Abstract{
Many real-world systems are profitably described as complex networks that grow over time. Preferential attachment and node fitness are two simple growth mechanisms that not only explain certain structural properties commonly observed in real-world systems, but are also tied to a number of applications in modeling and inference. While there are statistical packages for estimating various parametric forms of the preferential attachment function, there is no such package implementing non-parametric estimation procedures. The non-parametric approach to the estimation of the preferential attachment function allows for comparatively finer-grained investigations of the `rich-get-richer' phenomenon that could lead to novel insights in the search to explain certain nonstandard structural properties observed in real-world networks. This paper introduces the \proglang{R} package \pkg{PAFit}, which implements non-parametric procedures for estimating the preferential attachment function and node fitnesses in a growing network, as well as a number of functions for generating complex networks from these two mechanisms. The main computational part of the package is implemented in \proglang{C++} with \proglang{OpenMP} to ensure scalability to large-scale networks. In this paper, we first introduce the main functionalities of \pkg{PAFit} through simulated examples, and then use the package to analyze a collaboration network between scientists in the field of complex networks. The results indicate the joint presence of `rich-get-richer' and `fit-get-richer' phenomena in the collaboration network. The estimated attachment function is observed to be near-linear, which we interpret as meaning that the chance an author gets a new collaborator is proportional to their current number of collaborators. Furthermore, the estimated author fitnesses reveal a host of familiar faces from the complex networks community among the field's topmost fittest network scientists.
}
\Plainkeywords{temporal networks, dynamic networks, preferential attachment, fitness, rich-get-richer, fit-get-richer, R, C++, Rcpp, OpenMP} 

\Address{
  Thong Pham\\
  Mathematical Statistics Team \\ 
  RIKEN Center for Advanced Intelligence Project, Tokyo, Japan \\
  E-mail: \email{thong.pham@riken.jp}\\
}

\begin{document}


\section[Introduction]{Introduction}
\label{sec: sec_1_introduction}


  Since the end of the last century, complex networks have been increasingly used in modeling many temporal relations found in diverse fields~\citep{evolution_book, caldarelli_book, newman_book}. Some notable examples include collaboration networks between authors in a scientific field~\citep{newman2001clustering}, connection networks between computers on the Internet~\citep{barabasi-www}, and sexual relation networks between members of a community~\citep{sexual_net}. The primary motivation for using complex networks as a simplified representation of real-world systems is that they shed light on the behaviors of complex systems through the study of underlying patterns of connections. Although this is an over-simplification for systems depending heavily on domain-specific details, this approach nevertheless offers a first view of a system's topological properties, and can be used to guide subsequent in-depth analyses.

  Among the most important real-world network structural properties is degree distribution. Degree distribution lets us understand the proportion of highly and lowly connected nodes in a network. Since the highly connected nodes are key components of a network, this understanding in turn sheds light on the answers of important practical questions, including how to prevent the spreading of rumors~\citep{rumour_spreading}, how to stop a virus outbreak~\citep{epidemic_spreading}, and how to guard against cybernetic attacks~\citep{cybernatic}. 

 The degree distributions of many real-world networks have been found to be heavy-tailed~\citep{barabasi-albert}. The best-known heavy-tailed distribution in network science is the power-law, which is a distribution where the number of nodes in a network with degree $k$ is proportional to $k^{-\gamma}$ with $\gamma > 1$. Besides the power-law, there is emerging evidence that real-world network degree distributions have other heavy-tailed forms, including the log-normal~\citep{110year}, exponential~\citep{exponential}, stretched exponential~\citep{stretch_expo_cyberattack}, and power-law with exponential cut-off~\citep{clauset_cutoff}. 

  All of these heavy-tailed distributions differ from the light-tailed binomial degree distribution, which is characteristic of networks produced by the classical Erd{\"o}s-R{\'e}nyi (ER) random graph model~\citep{erdos-renyi}. This observation prompted network scientists to search for new modeling ingredients capable of explaining heavy-tailed degree distributions. What they found is that temporal complex network models incorporating growth mechanisms offer a powerful modeling framework for achieving this end. 

  Temporal complex network models, or temporal network models for short, are probabilistic generative models of real-world networks that change with time. In its most common form, a temporal network model assumes that a network grows gradually from some initial state by the addition of new nodes and edges over a large number of discrete time-steps. Some well-known basic models in the field of complex networks are the Barab{\'a}si-Albert (BA) model~\citep{barabasi-albert} and the Bianconi-Barab{\'a}si (BB) model~\citep{fitness}. More complex growth models that are used in the field include exponential random graph models~\citep{rsiena,tergm} and dynamic stochastic block models~\citep{dynsbm_paper}. Growth mechanisms, which govern how a node acquires new edges in the growth process, are the most important elements that distinguish different temporal network models.
 
This paper focuses on estimating two interpretable growth mechanisms: preferential attachment (PA) and node fitness. In the PA mechanism, the probability $P_i(t)$ that a node $v_i$ acquires a new edge at time $t$ is proportional to a positive function, $A_{k_i(t)}$, of its current degree $k_i(t)$. The function $A_{k}$ is called the attachment function. The name `preferential attachment' stems from the motivation for the mechanism: if $A_k$ is an increasing function, a highly connected node will acquire more edges than a lowly-connected node,  which is an appealing property in many real-world situations. From now, we will say that PA exists if $A_k$ is an increasing function. The opposite of PA, which we call anti-PA, occurs when $A_k$ is a decreasing function. Note, however, that the meaning we use here differs from the original meaning of the term `preferential attachment' used in the BA model, which means only the linear case of $A_k = k$. This linear form in fact has been long known in other fields with various names such as `rich-get-richer'~\citep{simon} and `cumulative advantage'~\citep{price1}. When  $A_k$ assumes the log-linear form of $k^\alpha$, with $\alpha$ called the attachment exponent, we have the generalized BA model~\citep{krapi}.

 In contrast with the PA mechanism, in the fitness mechanism the probability $P_i(t)$ that a node $v_i$ acquires a new edge depends only on the positive number $\eta_i$. The quantity $\eta_i$ is called the node fitness, or just fitness, of $v_i$ and can be interpreted as its intrinsic attractiveness. The fitness mechanism offers a simple way to express the variance in edge-acquisition potential between nodes of the same degree. For example, two early-career scientists with roughly the same number of collaborators at some point in time may acquire different numbers of collaborators in the future based on their intrinsic fitnesses. 
  
  The PA and node fitness mechanisms combine to produce a wide range of degree distributions. In their combined form, the probability $P_i(t)$ is proportional to the product of $A_{k_i(t)}$ and $\eta_i$:
\begin{equation}
P_i(t) \propto A_{k_i(t)} \times \eta_i. \label{eq: model}
\end{equation}
Based on the functional form of $A_k$ and the distribution of $\eta_i$'s, the model defined by Equation~\ref{eq: model} can produce networks with various degree distributions~\citep{fitness, fitness2,FirstToMarket,Kong}. In Section~\ref{sec: background_method}, we will discuss the relation of Equation~\ref{eq: model} with existing statistical models.

  Equation~\ref{eq: model} has a number of applications. Based on the functional forms of $A_k$ and $\eta_i$, we can test for the presence of one and/or the other of the `rich-get-richer' and `fit-get-richer' phenomena in a temporal network~\citep{pham3}. These two mechanisms have been advanced to explain another phenomenon called the `generalized friendship paradox'~\citep{friendship_paradox_original, friendship_paradox_generalized, friendship_paradox_model_1}. They are also used in inference problems in biological networks~\citep{paul-scalefree,timeline_adaptive_sampling}, the World Wide Web~\citep{Kong}, Internet topology
graphs~\citep{timeline_mcmc_icml}, and citation networks~\citep{barabasi_success_2013,barabasi_success_2016,Ronda-Pupo2018}. Finally, we can classify real-world temporal network data based on the estimated attachment exponent of $A_k$~\citep{kunegis}.

  While there are existing \proglang{R} packages that estimate PA in a growing network, including the packages \pkg{tergm}~\citep{tergm} and \pkg{RSiena}~\citep{rsiena}. These packages, however, rely on parametric methods to estimated the $A_k$ function. This means that one has to assume a functional form for $A_k$, rather than learning it from observed data without constraint. Non-parametric estimation of $A_k$ allows for a finer inspection of the `rich-get-richer' phenomenon~\citep{pham2,pham3}, and such methods have been used to provide clues to explain irregularities observed in real-world degree distributions~\citep{sheridan_paradox}.
 
  This paper introduces the \proglang{R} package \pkg{PAFit}~\citep{pafit}, which fills the gap with an implementation of the standard PA and node fitness non-parametric estimation procedures. In particular, we implement Jeong's method~\citep{jeong}, Newman's method~\citep{newman2001clustering} and the PAFit method~\citep{pham2,pham3} in the package. The first two are heuristic methods that are widely used to estimate non-parametrically the attachment function $A_k$ in isolation, while the last one is a statistical method that can non-parametrically estimate either $A_k$ (or $\eta_i$) in isolation or simultaneously estimate the two mechanisms. Although using PAFit is advisable in almost every circumstance, Jeong's method and Newman's method are still widely used and might still be appropriate in certain situations. Therefore, the inclusion of the two heuristic methods in the package is warranted. We discuss their strengths and shortcomings in Section~\ref{sec: background_method} when we provide an overview of the methodology and related statistical models. 

  The package also implements a variety of functions to simulate temporal networks from the PA and node fitness mechanisms, as well as functions to plot the estimated results and underlying uncertainties. We review \pkg{PAFit}'s main functions in Section~\ref{sec: background_package}. Before demonstrating their usages on three simulated examples in Section~\ref{sec: simulated_examples}, we discuss how \pkg{PAFit} relates with existing network analysis packages in Section~\ref{sec: relation_other_package}. We provide a systematic simulation to asses the results of the non-parametric joint estimation in Section~\ref{sec: simulation_study}, before showing a complete end-to-end work-flow analyzing a collaboration network of scientists from the field of complex networks in Section~\ref{sec: real_example}. Finally, concluding remarks are given in Section~\ref{sec: final}.
   
\section[Background]{Mathematical background}
\label{sec: background_method}
  Here we review the standard methods for estimating the attachment function $A_k$ and node fitnesses $\eta_i$ in a temporal network. In Section~\ref{sec: background_method_growth}, we state the network growth model used in the package as well as discuss its relation with exiting statistical models. We review the estimation of $A_k$ in isolation in Section~\ref{sec: background_method_onlypa}, then the estimation of the $\eta_i$'s in isolation in Section~\ref{sec: background_method_onlyfit}, and finally the joint estimation of $A_k$ and the $\eta_i$'s in Section~\ref{sec: background_method_joint}.

\subsection{Network model}
\label{sec: background_method_growth}
First we describe the General Temporal (GT) model~\citep{pham3} used in \pkg{PAFit}. The model is a generalization of many well-known temporal network models in the complex network field.
 
 Starting from some given initial graph $G_0$, the GT model generates a temporal network sequentially as follows: at time-step $t \geq 1$, the network $G_{t}$ is obtained by adding new edges and new nodes to $G_{t-1}$. The number of new edges and new nodes added at time $t$ is denoted as $m(t)$ and $n(t)$, respectively. The model assumes that the parameters governing the distributions of $G_{0}$, $m(t)$ and $n(t)$ do not involve $A_{k}$ and the $\eta_i$'s. Note that the term $k_i(t)$ is defined as the degree of node $v_i$ at the onset of time-step~$t$. On top of these structural preconditions, the GT model assumes that the probability that a node $v_i$ with degree $k_i(t)$ receives a new edge at time $t$ is given by the formula of Equation~\ref{eq: model}.

The GT model includes a handful of well-known growing network models, based on PA and node fitness as special cases; see Table~\ref{tab: real-models} for a summary. Unlike the BA or BB models, the GT model allows for the emergence of new edges between old nodes and can handle both undirect and directed networks. We refer readers to Supplementary Information Section S2.2 in~\cite{pham3} for the definition of the model in the case of undirected networks. The GT model is related to models used in the \proglang{R} packages \pkg{RSiena} and \pkg{tergm}. But we defer a discussion of these packages to Section~\ref{sec: relation_other_package}.

\begin{table}[!h]
\centering
\begin{tabular}{ l l l}
\hline
Temporal network model & Attachment function & Node fitness \\ \hline
Growing ER model~\citep{grow-random}& $A_k = 1$ & $\eta_i = 1$ \\ 
 BA model & $A_k = k$ & $\eta_i = 1$ \\ 
 Caldarelli model & $A_k = 1$ & Free \\    
 BB model & $A_k = k$ & Free \\  
\hline
\end{tabular}
\caption{Some well-known special cases of the GT model as defined by Equation~\ref{eq: model}. \label{tab: real-models}}
\end{table}

 Looking beyond the field of complex networks, the GT model bears some similarities to the contagious Poisson process~\citep{coleman,contagious_poisson} and the conditional frailty model~\citep{frailty_conditional_first,frailty_conditional}. In the contagious Poisson process, the initial propensity of each node plays a similar role to that of node fitness and the rate of enforcement represents the PA mechanism. In the conditional frailty model, while the frailty of each node describes the heterogeneity among nodes and thus is similar to node fitness, the event-based baseline hazard rate has the same effect as the non-parametric function~$A_k$.  

\subsection{Attachment function estimation}
\label{sec: background_method_onlypa}
 The methods for estimating the attachment function $A_k$ in isolation assume a simplified version of Equation~\ref{eq: model}, in which the $\eta_i$ are assumed to be $1$. Thus the probability $P_i(t)$ in Equation~\ref{eq: model} depends only on $A_k$. Perhaps the most frequently-encountered parametric version of this model is the log-linear form $A_k = k^\alpha$ with attachment exponent $\alpha$. Network scientists are particularly interested in estimating $\alpha$, since the asymptotic degree distribution of the network corresponds to simple regions of $\alpha$. If $\alpha$ is less than unity (the sub-linear case), then the degree distribution is a stretched exponential, while in the super-linear case of $\alpha > 1$, one node will eventually get all the incoming new edges~\citep{krapi}. It is only the linear case of $\alpha = 1$ that gives rise to a power-law distribution.
 
Concerning the above model, there are three main methods for estimating $A_k$: Jeong's method~\citep{jeong}, Newman's method~\citep{newman2001clustering}, and PAFit~\citep{pham2}. Jeong's method basically makes a histogram of the number of new edges $n_{k}$ connected to a node with degree $k$, then divides $n_k$ by the number of nodes with degree $k$ in the system to get~$A_k$~\citep{jeong}. Jeong's method has the merit of being simple, but estimates obtained using the method are subject to high variance and low accuracy~\citep{pham2}. By contrast, Newman's method combines a series of histograms for lower variance and higher accuracy~\citep{newman2001clustering}. Note that in PAFit we implemented a corrected version of Newman's original method~\citep{pham2}. The main drawback of  Newman's method is that the mathematical assumption behind its derivation holds only when $\alpha = 1$, thus the method amounts to an approximation when $\alpha \ne 1$~\citep{pham2}. 

The final method is PAFit~\citep{pham2}. It iteratively maximizes an objective function that is a combination of the log-likelihood of the model with a regularization term for $A_k$ by a Minorization-Maximization (MM) algorithm~\citep{MM}. There is a hyper-parameter, called $r$, in the method that controls the strength of the regularization. PAFit chooses $r$ automatically by cross-validation~\citep{pham3}. We defer the details to Section~\ref{sec: background_method_joint}. The method is not only able to recover $A_k$ accurately, but also can estimate the standard deviation of the estimated $A_k$ for each $k$~\citep{pham2}. Its main drawback is that it might be slow, since it is an iterative algorithm.
  
\subsection{Node fitness estimation} 
\label{sec: background_method_onlyfit}
When we consider only node fitnesses, there are two generative models in the literature with different assumptions regarding the functional form of $A_k$ in Equation~\ref{eq: model}. While the Caldarelli model~\citep{fitness2} assumes that $A_k$ is $1$ for all $k$, the BB model~\citep{fitness} assumes that $A_k = k$. Both models have been shown to generate networks with various heavy-tailed distributions~\citep{FirstToMarket,Kong}.

Node fitnesses in both models can be estimated by variants of the PAFit method proposed in~\cite{pham3}, by either setting $A_k = k$ for the BB model or $A_k = 1$ for the Caldarelli model. These estimation methods use MM algorithms that maximize the corresponding log-likelihood functions with a regularization term that regularizes the distribution of the $\eta_i$'s. More specifically, the inverse variance of this distribution is controlled by a hyper-parameter, called $s$, which is chosen automatically by cross-validation. We defer a more detail discussion to the next section. We note that node fitnesses in the BB model can also be estimated by the method in~\cite{Kong}. But since PAFit has been shown to outperform this method~\citep{pham3}, we did not include it in the package. 

\subsection{Joint estimation of the attachment function and node fitnesses} 
\label{sec: background_method_joint}
Finally, by using the full model in Equation~\ref{eq: model} the method PAFit in~\cite{pham3} can jointly estimate $A_k$ and $\eta_i$. We note this full model includes all the temporal network models shown in Table~\ref{tab: real-models}. For a more complete table, see Table 1 in \cite{pham3}.

The objective function of PAFit is a combination of the log-likelihood of the full model defined by Equation~\ref{eq: model} and two regularization terms: one for $A_k$ and one for $\eta_i$. While we refer readers to Supplementary Information Section S2.3 in~\cite{pham3} for a complete presentation, we will sketch here the log-likelihood function for the case of directed networks. Assume the set of observed snapshots is $\{G_t\}_{t = 0}^{T}$. Let $\mathbf{A} = [A_0\ A_1 \cdots A_{K-1}]^\top$ be the vector of the PA function and $\boldsymbol{\eta} = [\eta_1 \ \eta_2 \cdots \eta_{N}]$ be the vector of node fitnesses. Here $K$ is the maximum degree appearing in the growth process and $N$ is the total number of nodes at the end of the process. Let $z_{i}(t)$ be the number of new edges connected to node $v_i$ at time-step $t$. Equation~\ref{eq: model} implies that $\{z_{i}(t)\}_{i=1}^N$ follows a multinomial distribution with parameters $\{\pi_i(t)\}_{i=1}^N$, where

\begin{equation}
\pi_{i}(t) = \frac{A_{k_i(t)}\eta_i}{\sum_{j = 1}^N A_{k_{j}(t)}\eta_j}. \label{eq: multinomial_prob}
\end{equation}
Here we use the convention $k_{j}(t) = -1$ for a node that did not exist at time-step $t$ and $A_{-1} = 0$. Using Equation~\ref{eq: multinomial_prob}, one can write the likelihood of each snapshot $G_{1},\cdots, G_{T}$. The log-likelihood function of the temporal network $\{G_{t}\}_{t=0}^{T}$ is then the sum of the log-likelihood of each snapshot and is equivalent to:
\begin{equation}
l(\mathbf{A},\boldsymbol{\eta}) = \sum_{t=1}^{T}\sum_{i=1}^{N}z_{i}(t)\log A_{k_{i}(t)} + \sum_{t=1}^{T}\sum_{i=1}^{N}z_{i}(t)\log \eta_i - \sum_{t=1}^{T}\sum_{i=1}^{N}z_{i}(t)\log \sum_{j=1}^{N}A_{k_{j}(t)}\eta_j + C  , \label{eq: log_likelihood}
\end{equation}
with $C$ being the logarithm of the product of the probability mass functions of $G_0$, $m(t)$, and $n(t)$. Since the GT model, as stated in Section~\ref{sec: background_method_growth}, assumes that the parameters governing the distributions of $G_{0}$, $m(t)$ and $n(t)$ do not involve $A_{k}$ and $\eta_i$, we can treat $C$ as a constant.

The regularization term for $A_k$ is defined by
\begin{equation}
reg_A = - r\sum_{k = 1}^{K-2} w_k \left(\log A_{k+1} + \log A_{k-1} - 2 \log A_k \right)^2, \label{eq: A_regularize}
\end{equation}
with $r \ge 0$, $w_k = \sum_{t=1}^T m_k(t)$ and $m_k(t)$ the number of edges that connect to a degree $k$ node at time-step $t$.  
This term controls the shape of $A_k$. When $r$ is large, $A_k$ becomes more linear on a log-scale. In limiting as $r$ approaches infinity, we effectively assume that $A_k = k^\alpha$~\citep{pham3}. Thus this covers the case of $\alpha = 1$ in the BA model and the BB model, and the case of $\alpha = 0$ in the growing ER and the Caldarelli model. 

The regularization term for the node fitnesses is defined by
\begin{equation}
reg_F= \sum_{i=1}^N(\left(s-1\right)\log \eta_i - s \eta_i), \label{eq: eta_regularize}
\end{equation}
with $s > 0$. This term is the sum of the logarithms of Gamma distribution densities with mean~$1$ and variance $1/s$. The regularization is equivalent to placing such Gamma distributions as priors independently for each node fitness $\eta_i$. The larger the value of~$s$, the more tightly concentrated the values of $\eta_i$ become. If $s$ is infinitely large, then all $\eta_i$ will take the same value. This is equivalent to estimating the attachment function in isolation. 

To conclude: joint estimation with the above regularization terms also compasses the two cases of estimating either $A_k$ or $\eta_i$ in isolation. In particular, we maximize the following  objective function:
\begin{equation}
J(\mathbf{A},\boldsymbol{\eta}) = l(\mathbf{A},\boldsymbol{\eta}) + reg_A + reg_F, \nonumber
\end{equation}
with an MM algorithm. At each iteration, the algorithm replaces the objective function with an easier-to-maximize surrogate function and this surrogate function is maximized instead. The surrogate function is chosen in such a way that the objective function value is guaranteed to be nondecreasing over iterations. We refer the readers to~\cite{MM-tutorial} for the definition of a surrogate function and the techniques used to derive them. For a surrogate function, the variables are often separable, i.e., the partial derivative of one variable does not involve the others, and thus the maximization at each iteration, i.e., finding the variables by setting all the partial derivatives to zero, can be parallelized. While we refer readers to Supplementary Information Section S2.4 of~\cite{pham3} for a detailed discussion, the essence of the MM algorithms in \pkg{PAFit} is to linearize the term $\log \sum_{j=1}^NA_{k_{j}(t)}\eta_j$ in Equation~\ref{eq: log_likelihood} and to apply Jensen's inequality to make the variables in Equation~\ref{eq: A_regularize} separable.  

As mentioned in the two previous sections, the values of $r$ and $s$ are automatically selected by cross-validation. In particular, the dataset is divided into a learning part $\{G_t\}_{0}^{T_*}$ and a testing part $\{G_t\}_{T_*}^{T}$, where $T_*$ is the smallest positive number such that the ratio of the number of new edges in the learning part, i.e., $\sum_{t=1}^{T_*}\sum_{i=1}^{N}z_i(t)$, to that of the whole dataset, i.e., $\sum_{t=1}^{T}\sum_{i=1}^{N}z_i(t)$, is at least $p = 0.75$ (the default value). For each combination of $r$ and $s$, we use the learning data to get the estimated value of $\mathbf{A}$ and $\boldsymbol{\eta}$ and plug these estimated values into Equation~\ref{eq: log_likelihood} to calculate the log-likelihood of the testing data. The combination of $r$ and $s$ that maximize this log-likelihood is then chosen. The method then re-estimates $\mathbf{A}$ and~$\boldsymbol{\eta}$ using the whole dataset with the chosen combination of $r$ and $s$.

\section[overview_package]{Package overview}
\label{sec: background_package}
The \pkg{PAFit} package provides functions to simulate various temporal network models, gather essential network statistics from raw input data, and use these summarized statistics in the estimation of $A_k$ and the $\eta_i$ values. The heavy computational parts of the package are implemented in \proglang{C++} through the use of the \pkg{Rcpp} package~\citep{rcpp1,rcpp2,rcpp3}. Furthermore, multi-core machine users can enjoy a hassle-free speed up through \proglang{OpenMP} parallelization mechanisms implemented in the code. Apart from the main functions, the package also includes a real-world collaboration network dataset between scientists in the field of complex networks. Table~\ref{tab: summary_functions} summarizes the main functions in the package. In what follows, we will review the main package functions one by one. 

\begin{sidewaystable} [tbp] 
\centering
\begin{tabular}{ l l l}
\hline
Function & Main input & Output  \\  \hline

\code{generate_ER}  & network parameters & network from the growing ER model \\
\code{generate_BA} &  network parameters  &  network from the generalized directed BA model\\
\code{generate_BB}  & network parameters &  network from the BB model \\
\code{generate_fit_only}  &  network parameters & network from the Caldarelli  model\\ 
\code{generate_net}  & network parameters &  network from the GT model \\
\code{get_statistics}  & \code{PAFit_net} object containing the network & \code{PAFit_data} object containing summary statistics\\
\code{Jeong}  & \code{PAFit_net} object and \code{PAFit_data} object & estimated PA function by Jeong's method\\
\code{Newman}  & \code{PAFit_net} object and \code{PAFit_data} object & estimated PA function by Newman's method\\
\code{only_A_estimate}  & \code{PAFit_net} object and \code{PAFit_data} object & estimated PA function by PAFit\\
\code{only_F_estimate}  &  \code{PAFit_net} object and \code{PAFit_data} object & estimated node fitnesses by PAFit\\
\code{joint_estimate}  & \code{PAFit_net} object and \code{PAFit_data} object  & estimated PA function and node fitnesses by PAFit \\ 
\code{to_networkDynamic} & \code{PAFit_net} object & \code{networkDynamic} object \\
\code{from_networkDynamic} & \code{networkDynamic} object &  \code{PAFit_net} object \\
\code{to_igraph} & \code{PAFit_net} object & \code{igraph} object \\
\code{from_igraph} &  \code{igraph} object & \code{PAFit_net} object \\
\code{graph_to_file} & \code{PAFit_net} object & text file in either edge-list format or \code{gml} \\
\code{graph_from_file} & text file in edge-list format or \code{gml} & a \code{PAFit_net} object \\
\code{as.PAFit_net} & edge-list matrix & \code{PAFit_net} object \\
\code{test_linear_PA} & degree vector & \code{Linear_PA_test_result} object \\

\hline
\end{tabular}
\caption{Summary of the main functions in the \pkg{PAFit} package. \label{tab: summary_functions}}
\end{sidewaystable}

Firstly, most well-known temporal network models based on PA and node fitness mechanisms can be easily simulated using the package. \pkg{PAFit} implements \code{generate_BA} for the BA model, \code{generate_ER} for the growing ER model, \code{generate_BB} for the BB model, and \code{generate_fit_only} for the Caldarelli model. These functions have many customizable options. For example, the number of new edges at each time-step is a tunable stochastic variable; see Table~\ref{tab: summary_parameter_generating_functions} for descriptions of the parameters. They are actually wrappers of the more powerful \code{generate_net} function, which simulates networks with more flexible attachment function and node fitness settings. 

\begin{table} [!h] 
\centering
\begin{tabular}{ l l }
\hline
Parameter (default value) & Description \\  \hline
\code{N} (\code{1000})  & total number of nodes in the network\\
\code{num_seed} (\code{2}) & initial graph is a circle with \code{num_seed} nodes \\
\code{multiple_node} (\code{1})  & number of new nodes added at each time-step \\
\code{m} (\code{1})  & number of edges of a new node \\ 
\code{alpha} (\code{1}) & attachment exponent $\alpha$ when we assume $A_k = k^\alpha$ \\
\code{mode_f} (\code{"gamma"})  & distribution of node fitnesses: gamma, log-normal or power-law \\
\code{s} (\code{10})  & distribution of node fitnesses has mean $1$ and variance $1/s$ \\
\hline
\end{tabular}
\caption{Main parameters in network generating functions in the \pkg{PAFit} package. \label{tab: summary_parameter_generating_functions}}
\end{table}

  Each temporal network model generation function outputs a \code{PAFit_net} object, which is a list with four fields: \code{type}, \code{fitness}, \code{PA}, and \code{graph}. The \code{type} field is a string indicating the type of network: \code{"directed"} or \code{"undirected"}. This field is \code{"directed"} for the networks generated by the simulation functions. The \code{fitness} and \code{PA} fields contain the true node fitnesses and PA function, respectively. The \code{graph} field contains the generated temporal network in a three-column matrix format. Each row of this matrix is of the form \code{(id\_1 id\_2 time\_stamp)}. While \code{id\_1} and \code{id\_2} are IDs of the source node and the destination node, respectively, \code{time\_stamp} is the birth time of the edge. This is the so-called edge-list format in which raw temporal networks are stored in many on-line repositories~\citep{konect,snap}. We will discuss how to use functions provided by \pkg{PAFit} to convert this edge-list format to formats used in other network analysis packages in the next section. One can apply the function \code{plot} directly to a \code{PAFit\_net} object to visualize its contents. 

The second functionality of \pkg{PAFit} is implemented in \code{get\_statistics}. With its core part implemented in \proglang{C++}, this function efficiently collects all temporal network summary statistics that are needed in the subsequent estimation of PA and node fitnesses. The input network is assumed to be stored as a \code{PAFit\_net} object. One can use the function \code{graph\_from\_file} to read an edge-list graph from a text file into a \code{PAFit\_net} object, or convert an edge-list matrix to this class by the function \code{as.PAFit\_net}. 

The edge-list matrix is assumed to be in the same format as \pkg{PAFit} simulated graphs, i.e., each row is of the form \code{(id\_1 id\_2 time\_stamp)}. The node IDs are required to be integers greater than $-1$, but need not to be contiguous. Note that \code{(id -1 t)} describes a node \code{id} that appeared at time {t} without any edge. There are no assumptions on the values or data types of \code{time\_stamp}, other than that their chronological order is the same as what the \proglang{R} function \code{order} returns. Examples of timestamps that satisfy this requirement are the integer vector \code{1:T}, the format `yyyy-mm-dd', and the POSIX time.

The \code{get_statistics} function automatically handles both directed and undirected networks. It returns a list containing many statistics that can be used to characterize the network growth process. Notable fields are \code{m_tk} containing the number of new edges that connect to a degree-$k$ node at time-step~$t$, and \code{node_degree} containing the degree sequence, i.e., the degree of each node at each time-step.

The most important functionality of \pkg{PAFit} relates to the estimation of the attachment function and node fitnesses of a temporal network. This is implemented through various methods. There are three usages: estimation of the attachment function in isolation, estimation of node fitnesses in isolation, and the joint estimation of the attachment function and node fitnesses. 

The functions for estimating the attachment function in isolation are: \code{Jeong} for Jeong's method, \code{Newman} for Newman's method, and \code{only_A_estimate} for the PAFit method in \cite{pham2}. For estimation of node fitnesses in isolation, \code{only_F_estimate} implements a variant of the PAFit method in~\cite{pham3}. For the joint estimation of the attachment function and node fitnesses, we implement the full version of the PAFit method~\citep{pham3} in \code{joint_estimate}. The input of these functions is the output object of the function \code{get_statistics}. The output objects of these functions contain the estimation results as well as some additional information pertaining to the estimation process.  

In Table~\ref{tab: joint_estimate}, we show the input parameters of \code{joint_estimate}, the most important function in \pkg{PAFit}. This function takes the temporal network \code{net_object} and the summarized statistics \code{net_stat} as the main inputs. There are three parameters that control the estimation process: \code{p}, \code{stop_cond}, and \code{mode_reg_A}. The parameter \code{p} specifies the ratio of the number of new edges in the learning data to that of the full data in the cross-validation step. Following~\cite{pham3}, its default value is set at $0.75$. The parameter \code{stop_cond} specifies the threshold~$\epsilon$:~the iterative algorithm will continue until the relative difference in the objective function~$J(\mathbf{A},\boldsymbol{\eta})$ between two successive iterations falls below this threshold~\citep{pham3,quasi-Newton-MM}. The default value $\epsilon = 10^{-8}$ is set following~\cite{pham3}. The parameter \code{mode\_reg\_A} specifies the regularization term for $A_k$. The default value \code{mode\_reg\_A = 0} corresponds to the regularization term in~Equation~\ref{eq: A_regularize}~\citep{pham3}. When \code{mode\_reg\_A = 1}, the following regularization term is used: 
\begin{equation}
-r\sum_{k=2}^{K-1} w_k \left\{ \frac{\log{A_{k+1}} - \log{A_{k}}}{\log{(k+1)} - \log{k}} -\frac{\log{A_{k}} - \log{A_{k - 1}}}{\log{k} - \log{(k - 1)}} \right\}^2.
\end{equation}
Although this regularization term will enforce exactly the form $A_k = k^\alpha$, it is significantly slower to optimize this regularization term while the improvement over~Equation~\ref{eq: A_regularize} is little.

\begin{table} [!h] 
\centering
\begin{tabular}{ l l }
\hline
Parameter  & Default value \\  \hline
\code{net_object} & no default value \\
\code{net_stat} & \code{get_statistics(net\_object)} \\
\code{p}  &  $0.75$ \\
\code{stop_cond} & $10^{-8}$ \\
\code{mode_reg_A} & 0 \\
\hline
\end{tabular}
\caption{Parameters of the \code{joint\_estimate} function and their default values.\label{tab: joint_estimate}}
\end{table}

Finally, although one can roughly assess whether PA exists in the network by visual inspection of the estimated PA function,~\cite{handcock_test_linear_PA}  provide a method to test whether the linear PA-only case, i.e., $A_k = k$ and $\eta_i = 1$, is consistent with a given degree vector. We implemented this method in the function \code{test_linear_PA}. This function chooses the best fitted distribution to a given degree vector among a set of distributions by comparing the Akaike Information Criterion~(AIC)~\citep{aic} or the Bayesian Information Criterion~(BIC)~\citep{bic}. The set of distributions are Yule, Waring, Poisson, geometric, and negative binomial. The linear PA-only case corresponds to Yule or Waring~\citep{yule,waring}.

\section[relation]{Related network packages}
\label{sec: relation_other_package}

Since network analysis has been an important field for a long time, various aspects of it have been implemented in a large number of software packages. To our best effort, we have confirmed that the non-parametric joint estimation of PA and fitness mechanisms in a growing network is not implemented elsewhere. Restricting the discussion to packages in \proglang{R}, there are some notable implementations of related statistical network models. For example, stochastic block models in the packages \pkg{igraph}~\citep{igraph}, \pkg{sna}~\citep{sna}, \pkg{blockmodels}~\citep{blockmodels} and \pkg{dynsbm}~\citep{dynsbm_paper,dynsbm}; exponential random graph models in the packages \pkg{ergm}~\citep{ergm_2,ergm_1}, \pkg{tergm}~\citep{tergm}, \pkg{hergm}~\citep{hergm_article}, \pkg{btergm}~\citep{btergm}, and \pkg{RSiena}~\citep{rsiena}; and latent space models in the package \pkg{latentnet}~\citep{latentnet_2,latentnet_1}. 

The \pkg{dynsbm} package estimates a dynamic stochastic block model in which nodes are assumed to belong to some latent groups which can vary with time, and the edge weight between two nodes at any time follows some parametric distribution. The package can deal with both discrete and continuously weighted edges.

The \pkg{igraph} package contains the functions \code{sample_pa} and \code{sample_growing} which are the equivalents of \code{generate_BA} and \code{generate_ER} in \pkg{PAFit}, respectively. Although \pkg{igraph} also generates networks from many other mechanisms, it does not contain any function for estimating the PA function and/or node fitnesses. It does contain many functionalities for dealing with stochastic block models and various other network models.

Some of the above packages are included in the extensive meta-package \pkg{statnet}~\citep{statnet_1,statnet_2}. In \pkg{statnet}, packages that deal with temporal networks are: \pkg{networkDynamic}~\citep{network_dynamic}, \pkg{tsna}~\citep{tsna}, and \pkg{tergm}. The \pkg{networkDynamic} package provides the \code{networkDynamic} class to store dynamic networks and various functions to manipulate them. The \pkg{tsna} package calculates many temporal statistics of a dynamic network stored in a \code{networkDynamic} object. 

The closest packages to \pkg{PAFit} that estimate PA in a temporal network are \pkg{tergm} and \pkg{RSiena}, which implement sophisticated continuous-time and discrete-time Markov models. Regarding PA, all the implemented options in \pkg{tergm} and \pkg{RSiena} pertain to the parametric estimation of~$A_k$, in contrast to the non-parametric estimation methods implemented in \pkg{PAFit}. Although  it might be theoretically possible to describe a non-parametric $A_k$ function in \pkg{tergm} and \pkg{RSiena}, they contain no regularization terms for the joint estimation of the non-parametric PA function and node fitnesses. Joint estimation without regularization terms is very likely unable to recover the true parameters, since the number of parameters is typically high. On the other hand, \pkg{PAFit} is specifically designed for estimating $A_k$ non-parametrically with node fitnesses, since it has two regularization terms in Equations~\ref{eq: A_regularize} and~\ref{eq: eta_regularize}, together with the cross-validation step for selecting suitable regularization parameters.

\pkg{PAFit} provides functionalities to communicate with existing network analysis packages. Using \code{to_networkDynamic} and \code{from_networkDynamic}, one can convert a \code{PAFit_net} object to a \pkg{networkDynamic}'s \code{networkDynamic} object and vice versa. The functions \code{to_igraph} and \code{from_igraph} do the same for \pkg{igraph}'s \code{igraph} objects. The extensive functions of \pkg{statnet} and \pkg{igraph} packages can then be used. One can also output the graph stored in a \code{PAFit_net} object to the universal \proglang{gml} format by the function \code{graph_to_file}, or read from a \proglang{gml} file by the function \code{graph_from_file}. 
 
\section{Package usage}
\label{sec: simulated_examples}

Here we show three usages of \pkg{PAFit}: the estimation of the attachment function $A_k$ in isolation in Section~\ref{sec: sec_4_2_pa_isolation}, the estimation of node fitnesses $\eta_i$ in isolation in Section~\ref{sec: sec_4_2_only_fitness}, and the joint estimation of $A_k$ and the $\eta_i$ values in Section~\ref{sec: sec_4_3_joint}.

\subsection{Attachment function estimation}
\label{sec: sec_4_2_pa_isolation}
First we generate a network from a directed version of the BA model, called Price's model~\citep{price1}. From the initial graph with two nodes and one edge, one new node with $m = 5$ new edges is added at each time-step until the number of nodes is $N = 1000$.
\begin{Sinput}
R> set.seed(1)
R> library("PAFit")
R> sim_net_1 <- generate_BA(N = 1000, m = 5)
\end{Sinput} 
Recall that $A_k$ is linear in the BA model, i.e., the attachment exponent $\alpha$ is equal to $1$, and the node fitnesses are uniform. 

\begin{figure}[!h]
\centering
\begin{subfigure}{0.325\textwidth}
\centering
\includegraphics[width=\linewidth]{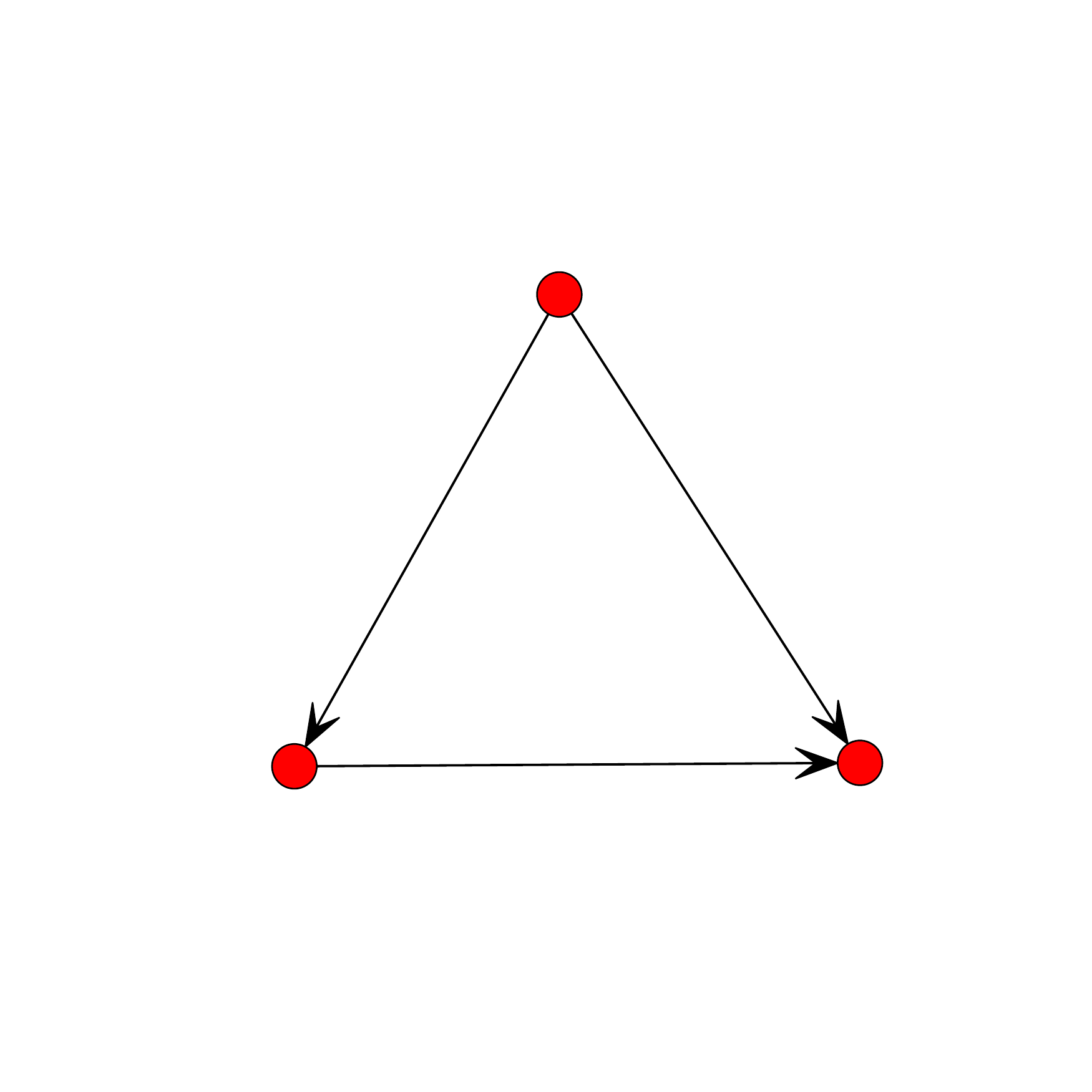}
\caption{Snapshot at $t = 1$.\label{fig : plot_simulated_1}}
\end{subfigure}
\begin{subfigure}{0.325\textwidth}
\centering
\includegraphics[width=\linewidth]{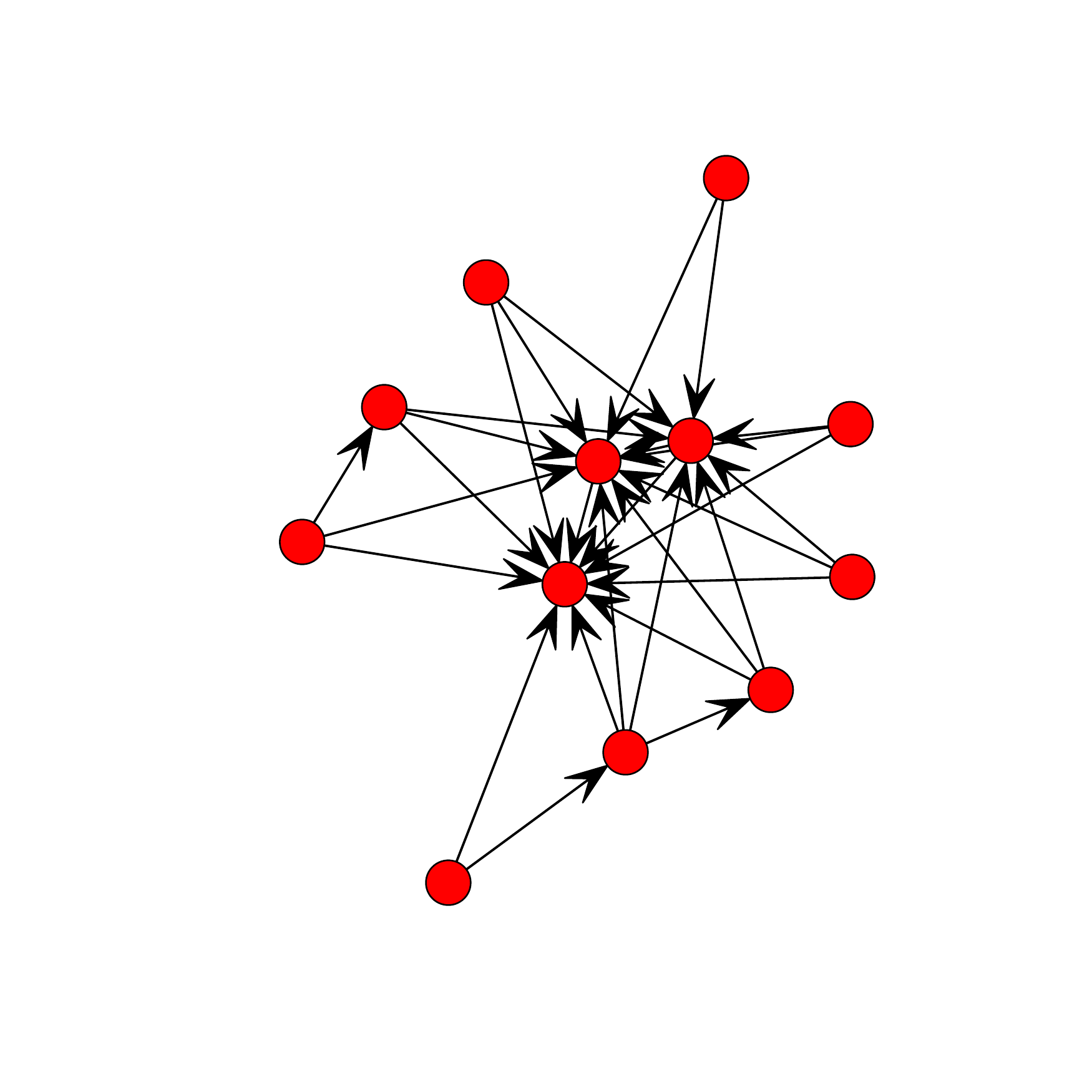}
\caption{Snapshot at $t = 10$.\label{fig : plot_simulated_2}}
\end{subfigure}
\begin{subfigure}{0.325\textwidth}
\centering
\includegraphics[width=\linewidth]{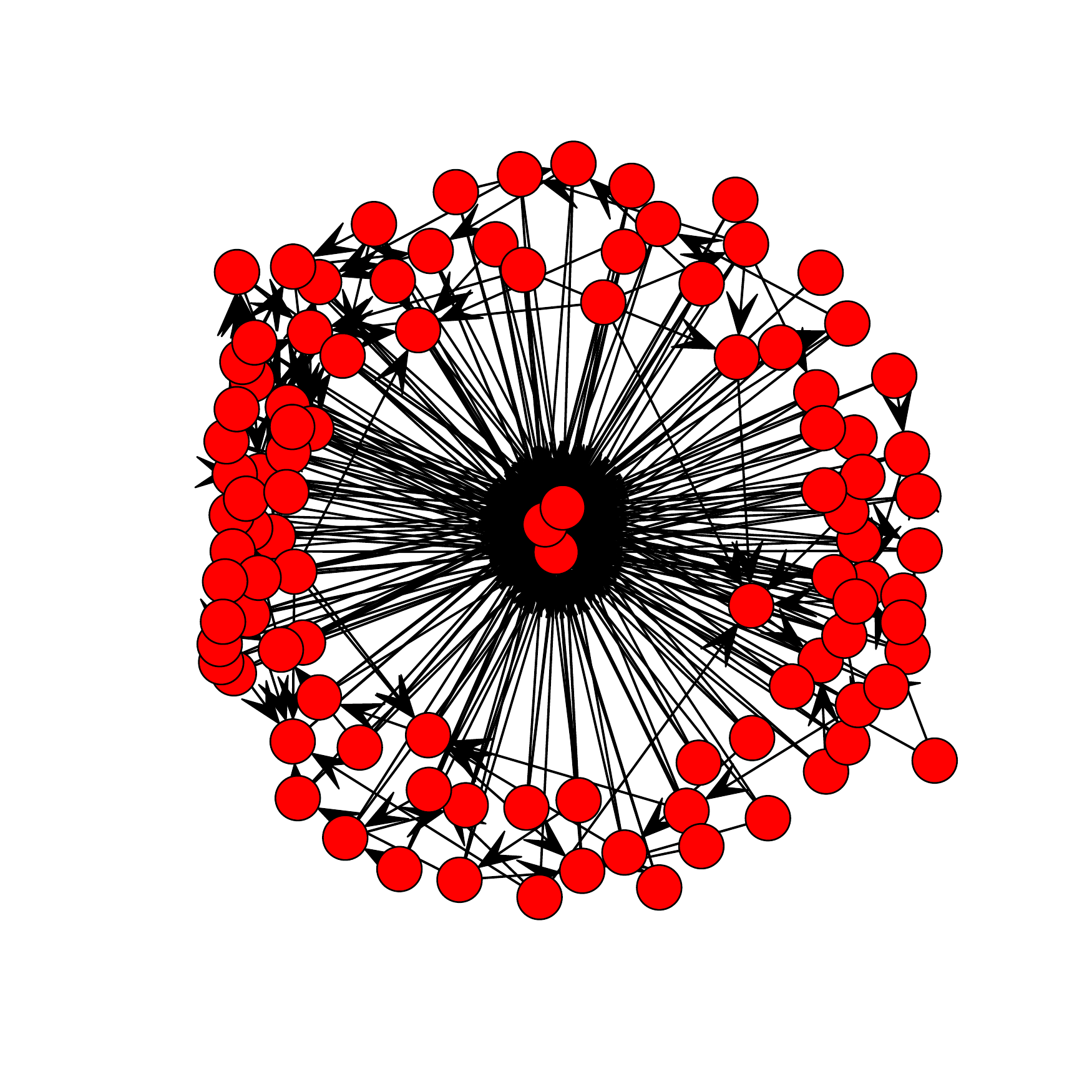}
\caption{Snapshot at $t = 100$.\label{fig : plot_simulated_3}}
\end{subfigure}

\begin{subfigure}{0.325\textwidth}
\centering
\includegraphics[width=\linewidth]{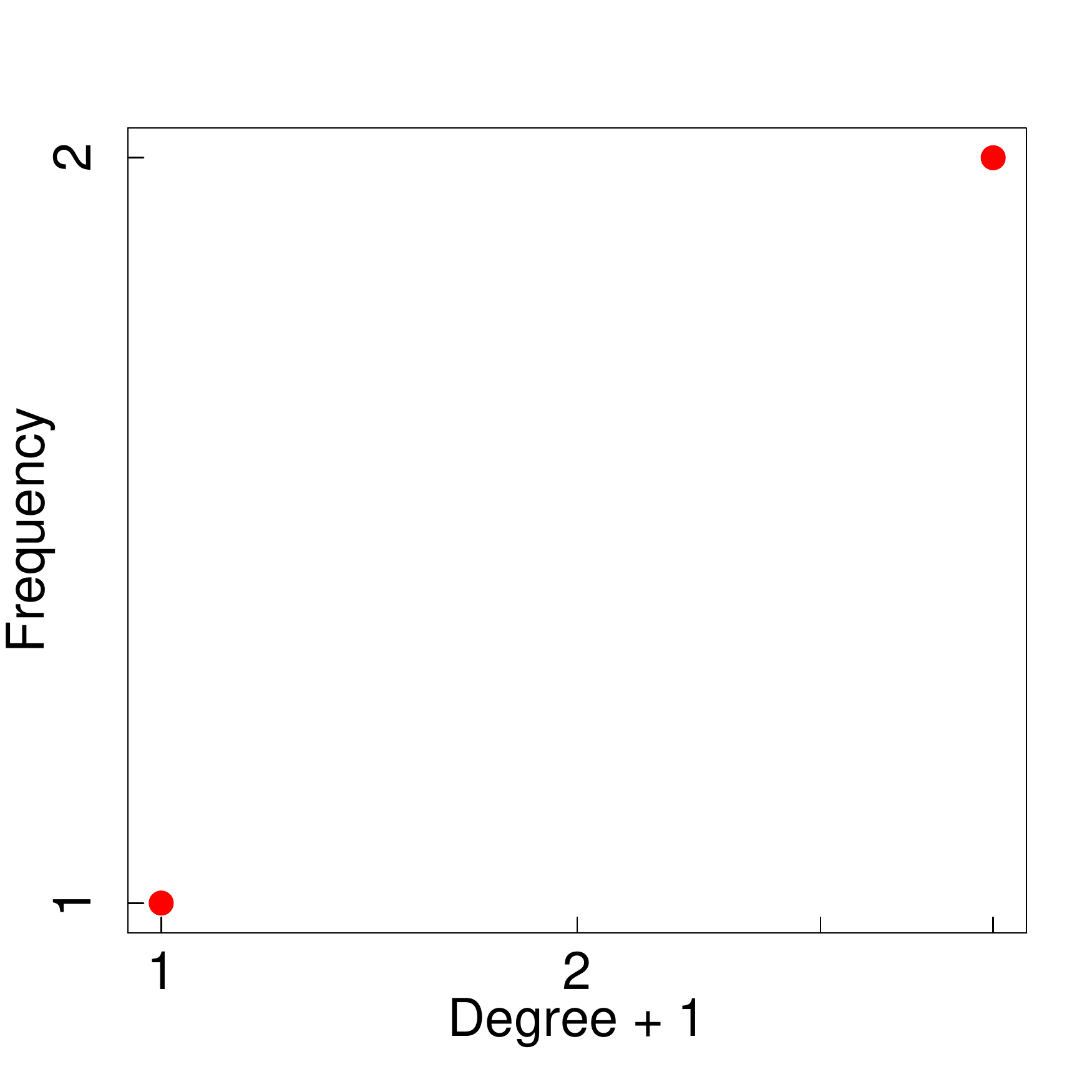}
\caption{Degree distribution at $t = 1$.\label{fig : plot_simulated_1_deg}}
\end{subfigure}
\begin{subfigure}{0.325\textwidth}
\centering
\includegraphics[width=\linewidth]{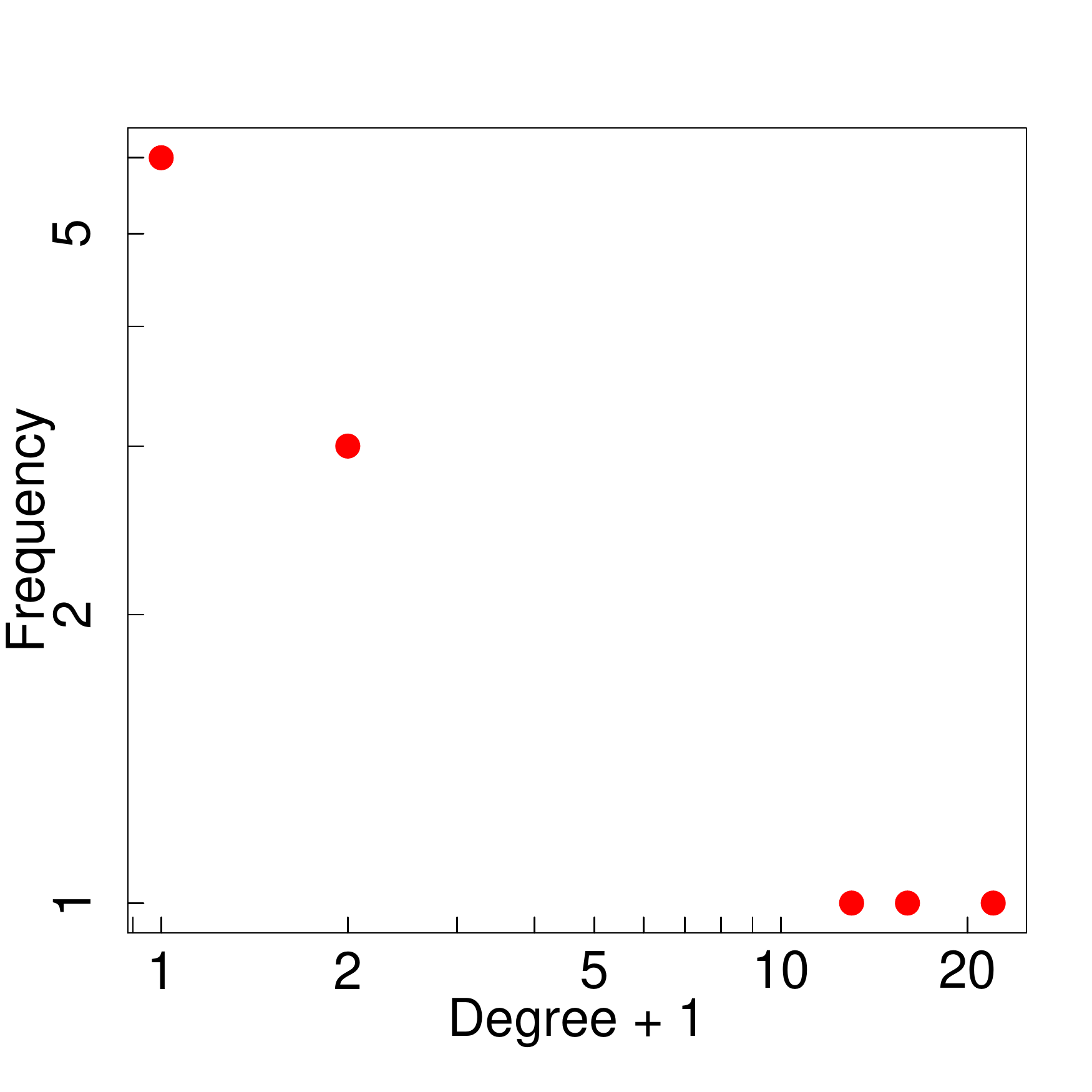}
\caption{Degree distribution at $t = 10$.\label{fig : plot_simulated_2_deg}}
\end{subfigure}
\begin{subfigure}{0.325\textwidth}
\centering
\includegraphics[width=\linewidth]{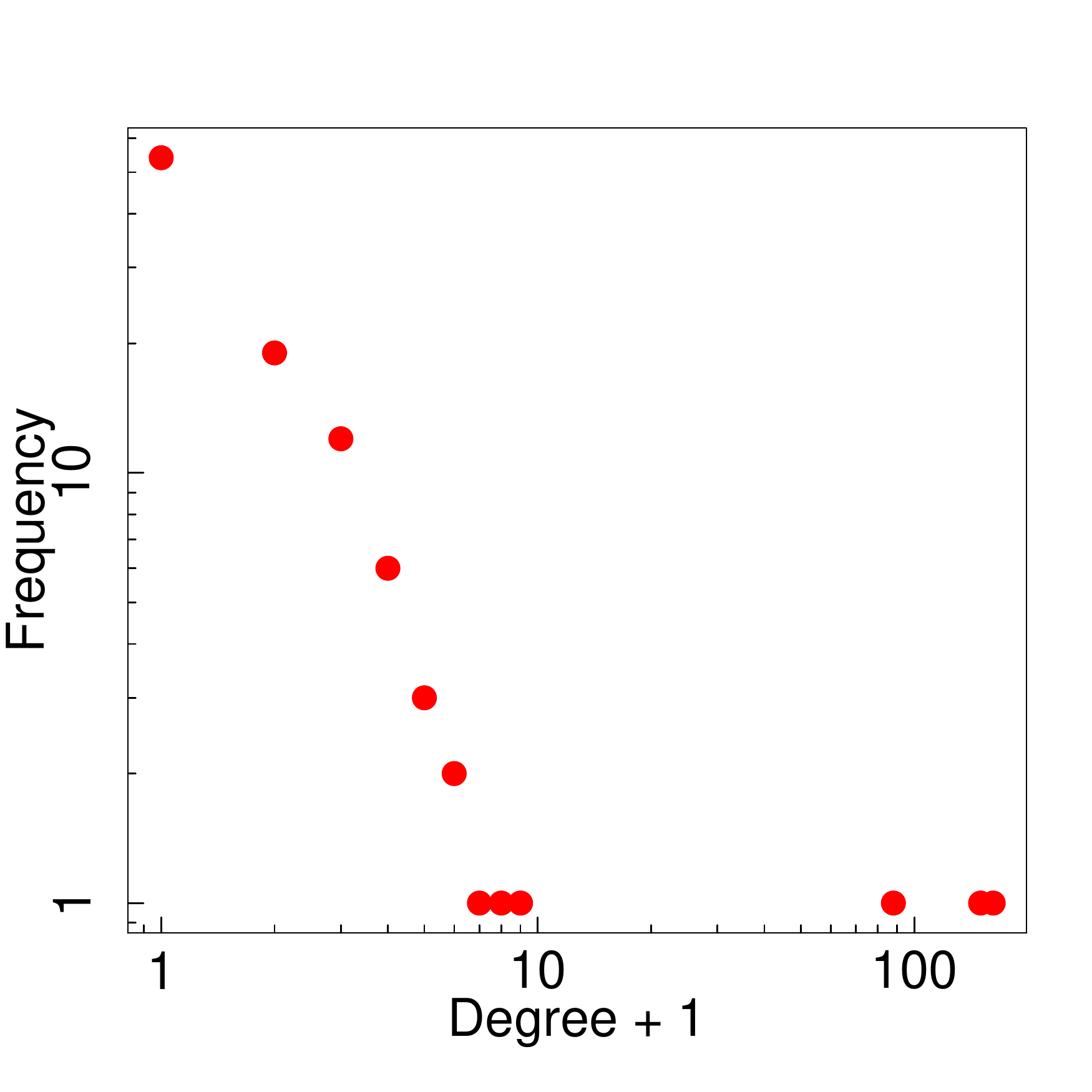}
\caption{Degree distribution at $t = 100$.\label{fig : plot_simulated_3_deg}}
\end{subfigure}
\caption{Network snapshots and their corresponding in-degree distributions at time-steps $t = 1$, $10$, and $100$. The temporal network, \code{sim\_net\_1}, is generated from Price's model with total number of nodes $N= 1000$. \label{fig : plot_simulated}}
\end{figure}

One can observe the emergence of hubs in this network by visualizing the generated graph at various time-steps by the function \code{plot}. The following script plots the network snapshot at time $t = 1$ in Figure~\ref{fig : plot_simulated_1} and its corresponding degree distribution in Figure~\ref{fig : plot_simulated_1_deg}:
\begin{Sinput}
R> plot(sim_net_1, slice = 1, arrowhead.cex = 3, vertex.cex = 3)
R> plot(sim_net_1, slice = 1, plot = "degree", cex = 3, cex.axis = 2,
+    cex.lab = 2)
\end{Sinput} 
Note that if the network is directed, as it is in this example, the option \code{plot = "degree"} will plot the in-degree distribution. In the same way, we plot network snapshots at time $t = 10$ and $t = 100$ in Figures~\ref{fig : plot_simulated_2} and~\ref{fig : plot_simulated_3} and their corresponding degree distributions in Figures~\ref{fig : plot_simulated_2_deg} and~\ref{fig : plot_simulated_3_deg}.

The next step is to use the function \code{get_statistics} to get the summary statistics for the temporal network:
\begin{Sinput}
R> stats_1 <- get_statistics(sim_net_1)
\end{Sinput} 
With \code{stats_1} containing all the needed summary statistics, we then apply the three methods of estimating the attachment function in isolation: 
\begin{Sinput} 
R> result_Jeong <- Jeong(sim_net_1, stats_1)          
R> result_Newman <- Newman(sim_net_1, stats_1)          
R> result_PA_only <- only_A_estimate(sim_net_1, stats_1) 
\end{Sinput}
Let us explain \code{result_PA_only} in more detail. Information on the estimated results as well as the estimation process can be viewed by invoking \code{summary}:
\begin{CodeChunk} 
\begin{CodeInput}
R> summary(result_PA_only)
\end{CodeInput}
\begin{CodeOutput}
Estimation results by the PAFit method. 
Mode: Only the attachment function was estimated. 
Selected r parameter: 0.1 
Estimated attachment exponent: 1.001139 
Attachment exponent ± 2 s.d.: (0.9908913,1.011387)
-------------------------------------------
Additional information: 
Number of bins: 50 
Number of iterations: 63 
Stopping condition: 1e-08
\end{CodeOutput}
\end{CodeChunk}
As stated in Section~\ref{sec: background_method}, the PAFit method first finds the $r$ parameter, which regularizes the PA function, by cross-validation, and then estimates $A_k$ using the chosen $r$. The estimated function can be accessed via \code{\$estimate_result\$k} and \code{\$estimate_result\$A} of \code{result_PA_only}. From this estimated function, the attachment exponent $\alpha$ (when we assume $A_k = k^\alpha$) and its standard deviation are also estimated. Here $\hat{\alpha}$ is $1.001 \pm 0.01$ as we can see from the output of \code{summary}. These values can be accessed via \code{\$estimate_result\$alpha} and \code{\$estimate_result\$ci}.

 The output also reveals that \pkg{PAFit} applies binning with $50$ bins by default. In this procedure, we divide the range of $k$ into bins consisting of consecutive degrees, and assume that all $k$ in a bin have the same value of $A_k$. Binning is an important regularization technique that significantly stabilizes the estimation of the attachment function~\citep{pham2}. In this example, the center of each bin is stored in the field \code{\$center\_k} of \code{stats\_1}.

\begin{figure}
\centering
\begin{subfigure}{0.32\textwidth}
\centering
\includegraphics[width=\linewidth]{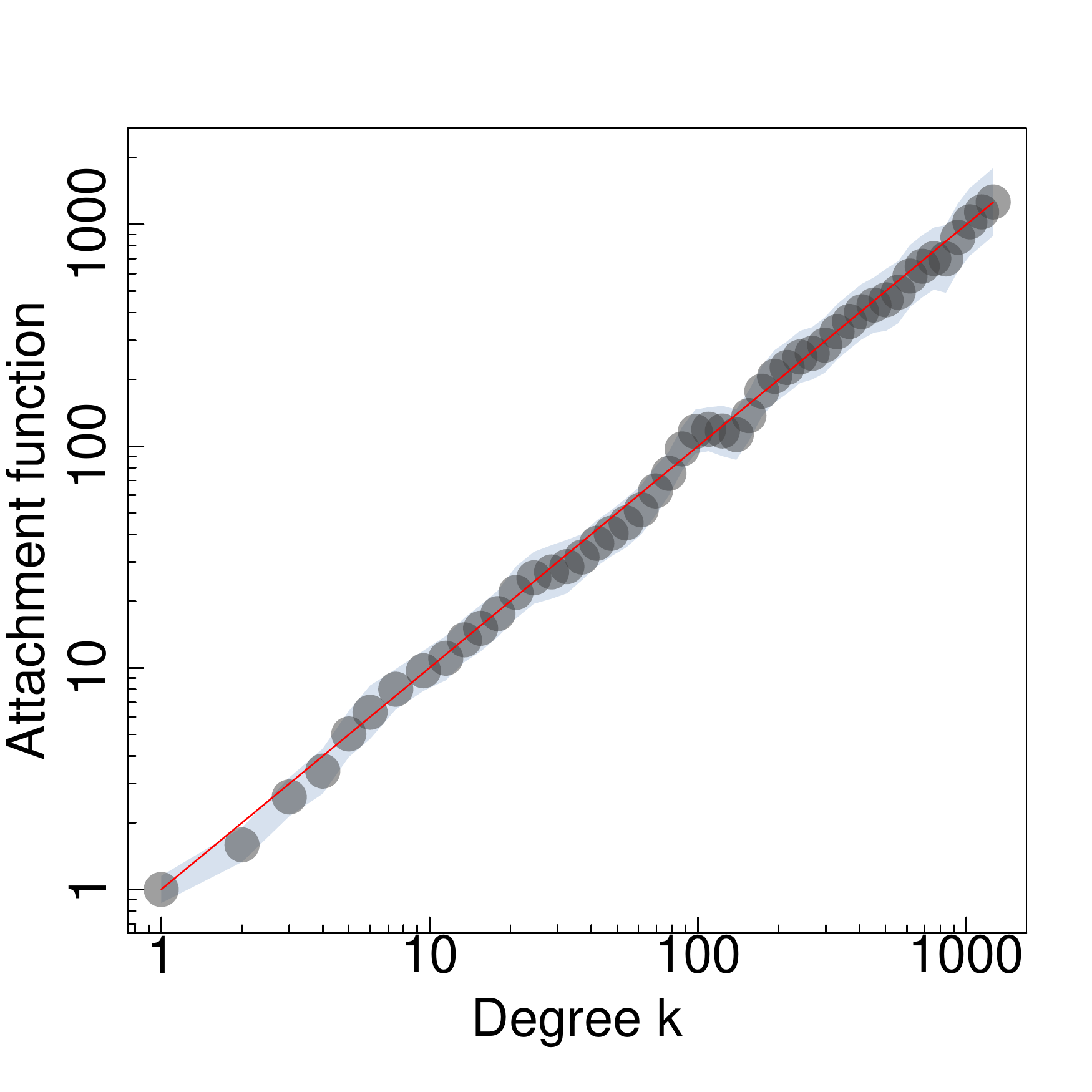}
\caption{PAFit \\ ($\hat{\alpha} = 1.001 \pm 0.01$)\label{fig: pa_only_example_PAFit}}
\end{subfigure}
\begin{subfigure}{0.32\textwidth}
\centering
\includegraphics[width=\linewidth]{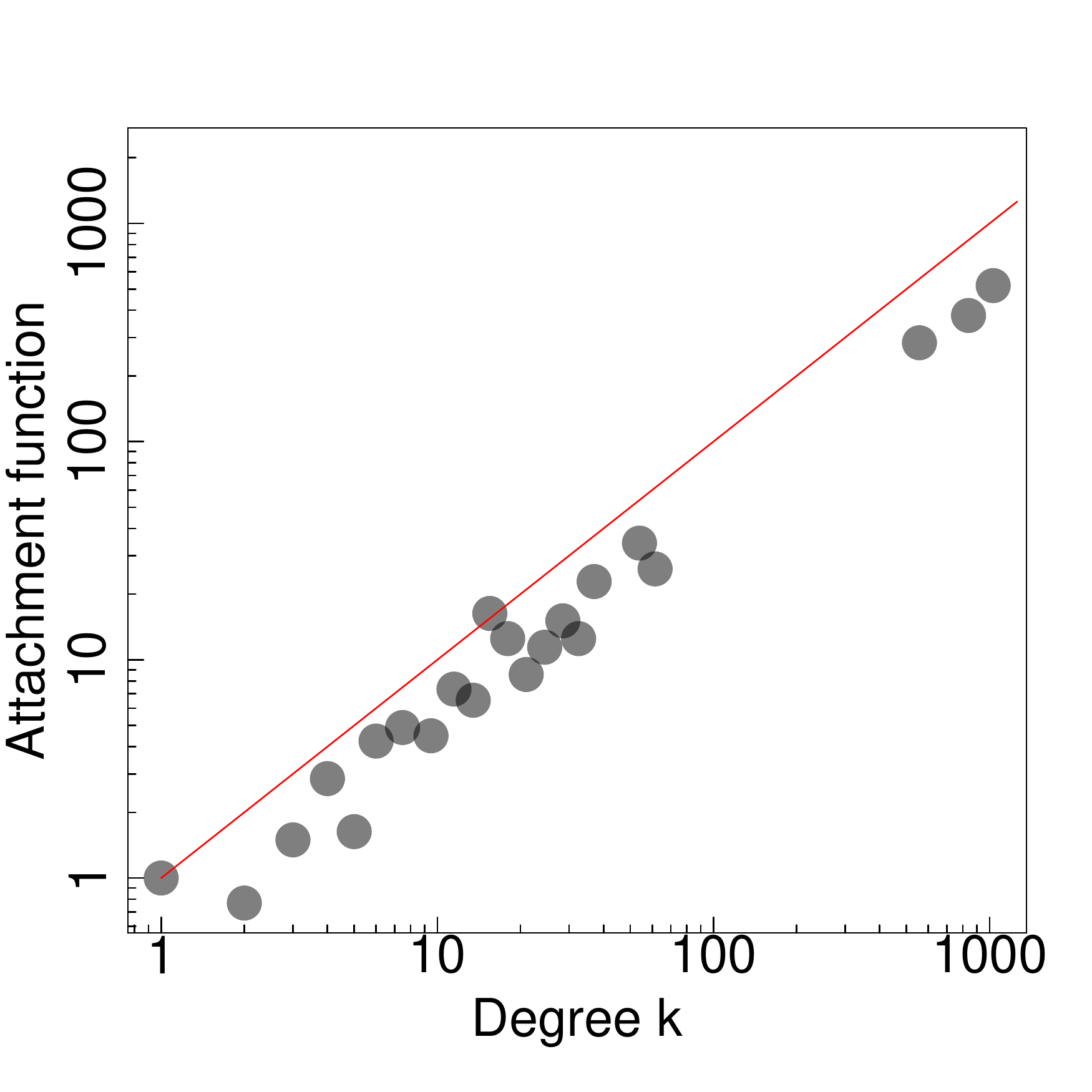}
\caption{Jeong's method \\ ($\hat{\alpha} = 0.96 \pm 0.07$)\label{fig: pa_only_example_Jeong}}
\end{subfigure}
\begin{subfigure}{0.32\textwidth}
\centering
\includegraphics[width=\linewidth]{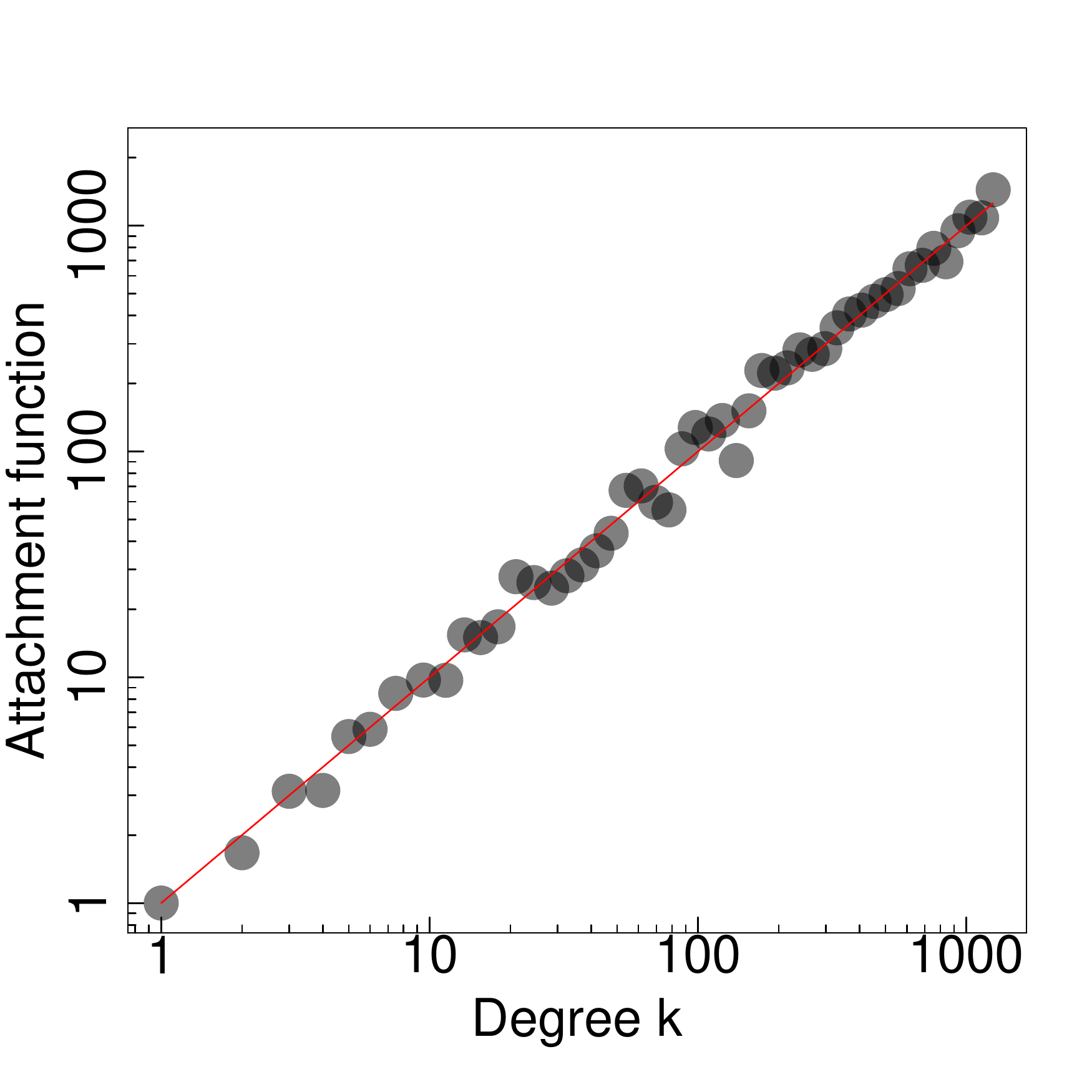}
\caption{Newman's method \\ ($\hat{\alpha} = 1.01 \pm 0.02$)\label{fig: pa_only_example_Newman}}
\end{subfigure}
\caption{Estimating the attachment function in isolation from \code{sim\_net\_1}. The true attachment function is $A_k = k^\alpha$ with attachment exponent $\alpha = 1$. We also show the estimated $\alpha$ and the interval of the estimated $\alpha\ \pm$ 2 s.d. provided by each method. \label{fig: pa_only_example}}
\end{figure}

Since the center of a bin is also the PA value corresponding to that bin in the linear PA case, we can plot the estimated attachment function together with the true attachment function using the following script, which produces the plot of Figure~\ref{fig: pa_only_example_PAFit}. The options \code{min\_A} and \code{max\_A} specify the minimum and maximum values in the vertical axis of the plot, respectively.
\begin{Sinput}
R> plot(result_PA_only,stats_1, min_A = 1, max_A = 2000, 
+    cex = 3, cex.axis = 2, cex.lab = 2)
R> lines(stats_1$center_k, stats_1$center_k, col = "red")
\end{Sinput}
The estimation results of Jeong's method and Newman's method can be plotted in a similar way, and are shown in Figures~\ref{fig: pa_only_example_Jeong} and~\ref{fig: pa_only_example_Newman}, respectively.

Overall, Newman's method and PAFit estimate the attachment function $A_k$ about equally well, while Jeong's method is found to underestimate the function and also exhibits high variance. This can also be observed in the estimated attachment exponent of the three methods: Newman's method and PAFit recover the true $\alpha$, while Jeong's method underestimates it. Note that in PAFit we also obtain the interval of the estimated $A_k$ $\pm$ 2 s.d. (lightblue region in Figure~\ref{fig: pa_only_example_PAFit}), which are unavailable in the other two methods. This is a significant advantage of PAFit over the other two methods since it allows the user to quantify uncertainties in the result.

\subsection{Node fitnesses estimation}
\label{sec: sec_4_2_only_fitness}
 Here we estimate node fitnesses from a BB model generated network with the assumption that $A_k = k$. To demonstrate the functionality of the package, we generate a BB network with a nonstandard setting: 
\begin{Sinput} 
R> sim_net_2 <- generate_BB(N = 1000, num_seed = 100, multiple_node = 100, 
+    m = 15, s = 10) 
\end{Sinput}
This network grows from a seed network with $N_0 = 100$ nodes where the nodes form a line graph. The value of $N_0$ can be specified by \code{num_seed}. At each time-step we add $n = 100$ new nodes where each node has $m = 15$ new edges. The values of $n$ and $m$ can be specified via \code{multiple_node} and \code{m}, respectively. The total number of nodes in the final network is $N = 1000$. Finally,  the distribution from which we generate node fitnesses is the Gamma distribution with mean $1$ and inverse variance $s = 10$.

\begin{figure}
\centering
\begin{subfigure}{0.49\textwidth}
\centering
\includegraphics[width=\textwidth]{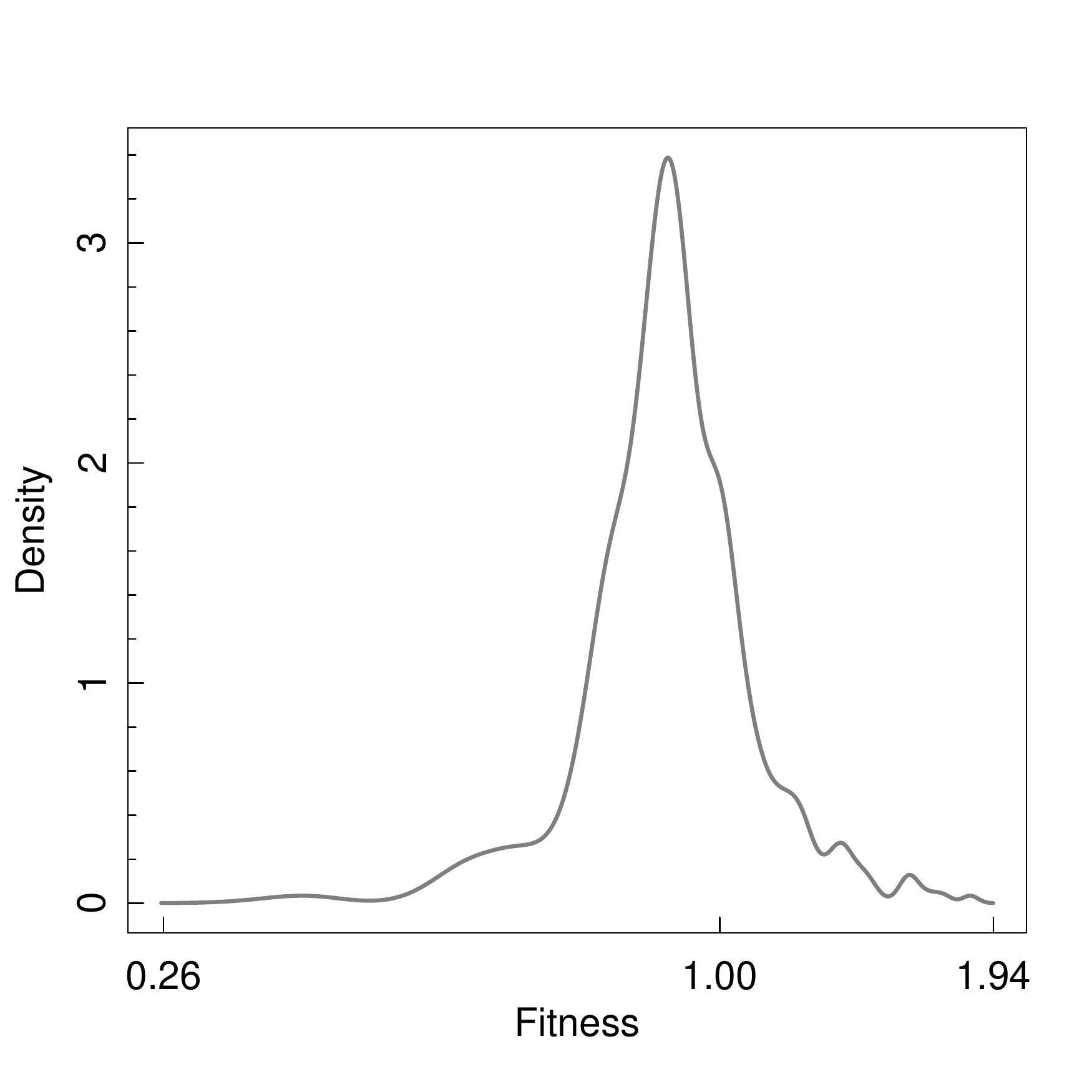}
\caption{Distribution of estimated fitnesses.\label{fig: fit_only_dist}}
\end{subfigure}
\begin{subfigure}{0.49\textwidth}
\centering
\includegraphics[width=\textwidth]{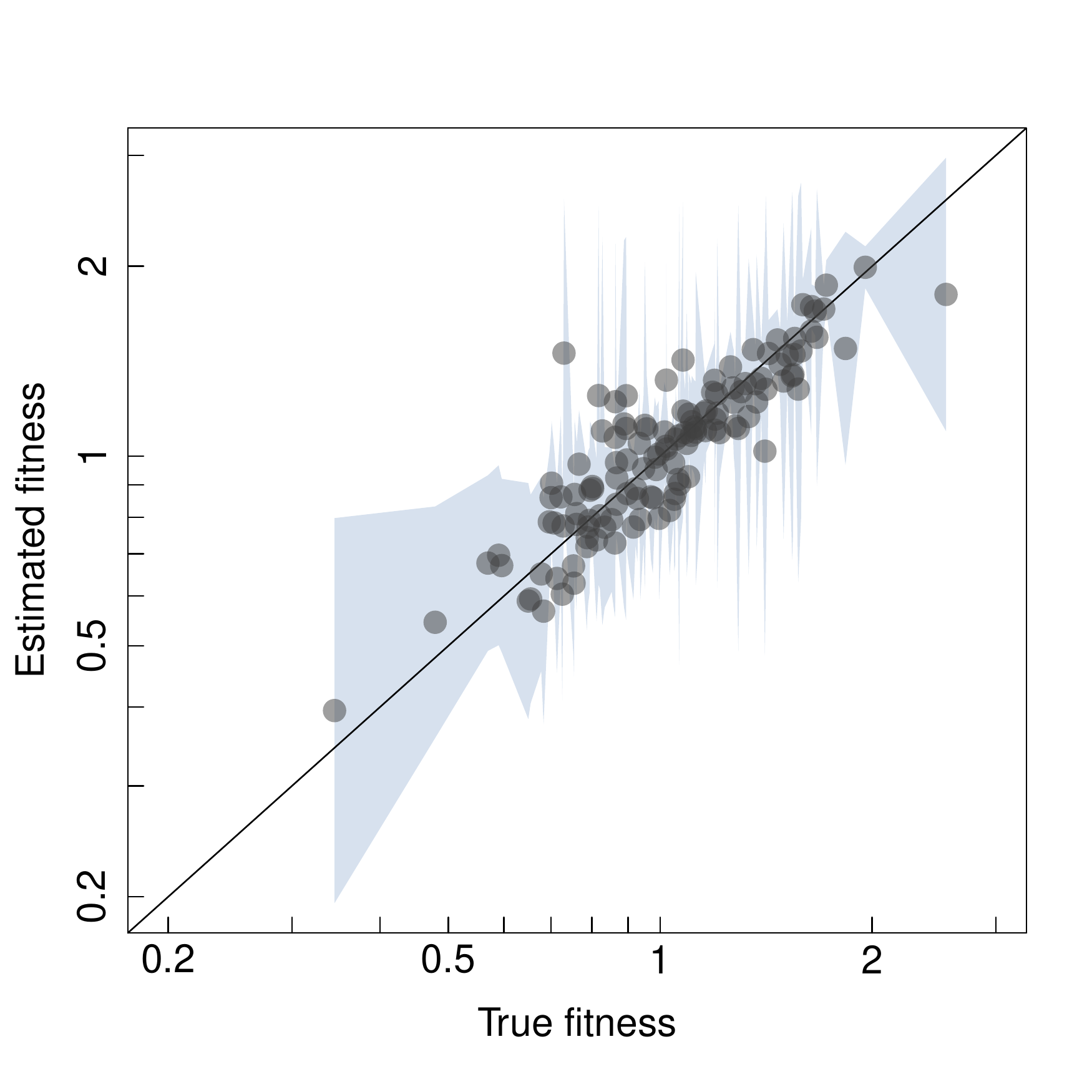}
\caption{Estimated fitnesses versus true fitnesses.\label{fig: fit_only_versus_true}}
\end{subfigure}
\caption{Estimating node fitnesses in isolation from \code{sim\_net\_2}, which is generated with attachment function $A_k = k$. The true node fitnesses are sampled from a Gamma distribution with mean $1$ and inverse variance $10$. The attachment function in the estimation method is fixed at $A_k = k$. In panel b, we only plot nodes for which the number of acquired new edges is at least $5$. \label{fig: fit_only_example}}
\end{figure}

Next we get the network summary statistics and then apply the estimation function:
\begin{Sinput}
R> stats_2 <- get_statistics(sim_net_2)
R> result_fit_only <- only_F_estimate(sim_net_2, stats_2)
R> plot(result_fit_only, stats_2, plot = "f",
+    cex = 2, cex.axis = 1.5, cex.lab = 1.5) 
\end{Sinput}
The final line of the snippet generates the distribution of estimated node fitnesses shown in Figure~\ref{fig: fit_only_dist}. 

The function \code{only_F_estimate} estimates node fitnesses under the assumption that $A_k = k$ by default. But one also can estimate node fitnesses in the Caldarelli model, i.e., $A_k = 1$ for all $k$, with the option \code{model_A = "Constant"}. The function \code{only_F_estimate} works by first finding the estimated value $\hat{s}$ of $s$ by cross-validation, and then using $\hat{s}$ in the subsequent estimation of node fitnesses. The summary information of the estimation result can be viewed by invoking \code{summary}:
\begin{CodeChunk} 
\begin{CodeInput}
R> summary(result_fit_only)
\end{CodeInput}
\begin{CodeOutput}
Estimation results by the PAFit method. 
Mode: Only node fitnesses were estimated. 
Selected s parameter: 8 
-------------------------------------------
Additional information: 
Number of bins: 50 
Number of iterations: 19 
Stopping condition: 0.00000001 
\end{CodeOutput}
\end{CodeChunk}
The method slightly under-estimated $s$. We can check whether the node fitnesses were estimated well by plotting the estimated fitnesses versus the true fitnesses by running the following command:
\begin{Sinput}
R> plot(result_fit_only, stats_2, true_f = sim_net_2$fitness, 
+    plot = "true_f", cex = 2, cex.axis = 1.5, cex.lab = 1.5)
\end{Sinput}

This will produce the plot of Figure~\ref{fig: fit_only_versus_true}. It turns out that the estimated node fitnesses agree pretty well with the true node fitnesses. We note that the light blue band around the $\hat{\eta_i}$ values depicts the intervals of $\hat{\eta_i}$ $\pm$ 2 s.d.. The upper and lower values can be accessed via \code{$estimate_result$upper_f} and \code{$estimate_result$lower_f} of \code{result_fit_only}, respectively. 

\subsection{Joint estimation of the attachment function and node fitnesses}
\label{sec: sec_4_3_joint}

Here we show how to estimate the attachment function and node fitnesses simultaneously. We need to assume in Section~\ref{sec: sec_4_2_pa_isolation} the equality of all $\eta_i$ for the estimation of $A_k$ in isolation, and in Section~\ref{sec: sec_4_2_only_fitness} a specific functional form of $A_k$ for the estimation of $\eta_i$ in isolation. Such assumptions become unnecessary when we perform joint estimation, since the appropriate functional forms will be automatically enforced through the regularization parameters $r$ and~$s$, which will be chosen by cross-validation. We recommend the joint estimation procedure as the standard estimation procedure in this package, unless there is a specific reason to justify the one or the other of these assumptions.

\begin{figure}
\centering
\begin{subfigure}{0.32\textwidth}
\centering
\includegraphics[width=\textwidth]{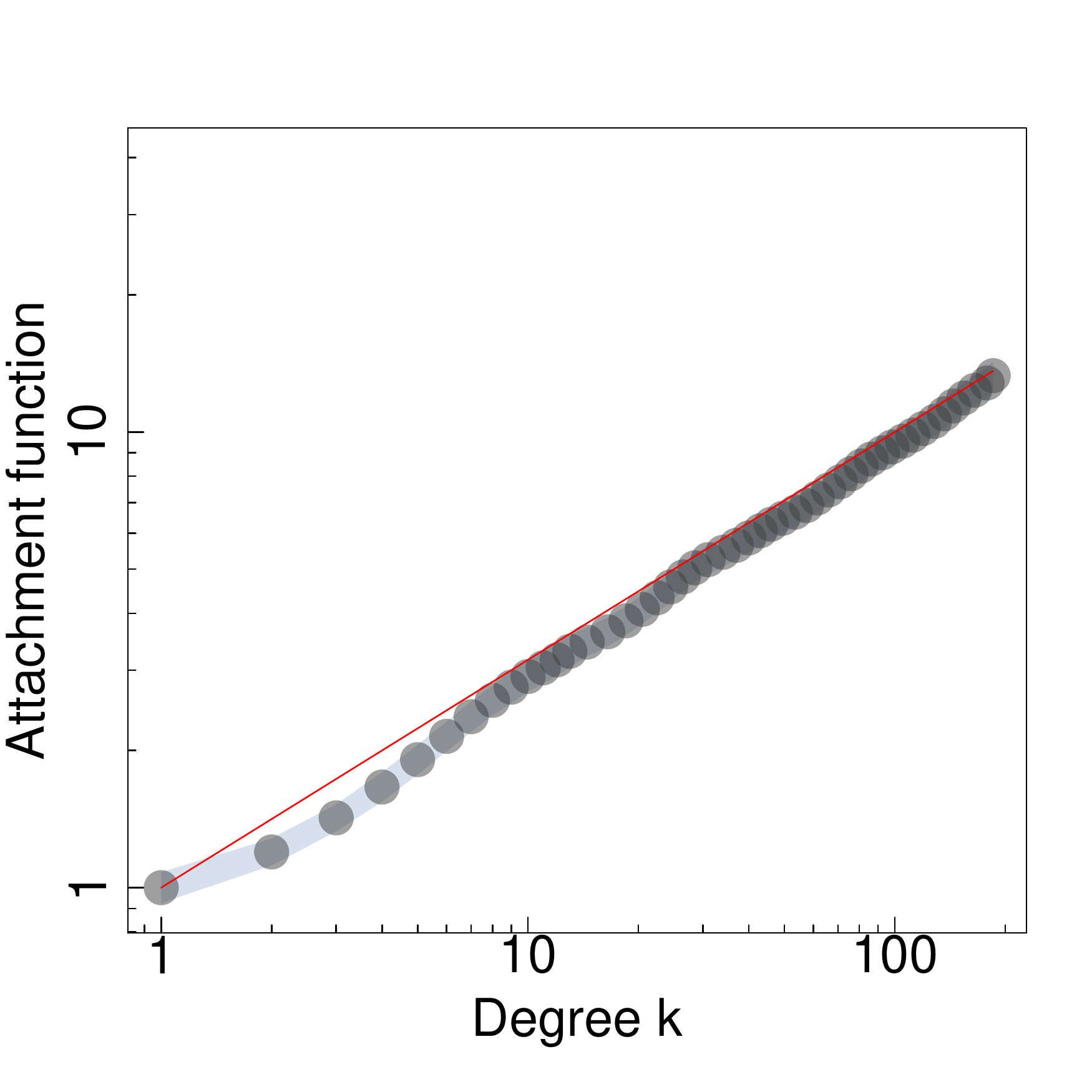}
\caption{Estimated attachment function ($\hat{\alpha} = 0.52 \pm 0.01$).} \label{fig: pafit_pa}
\end{subfigure}
\begin{subfigure}{0.32\textwidth}
\centering
\includegraphics[width=\textwidth]{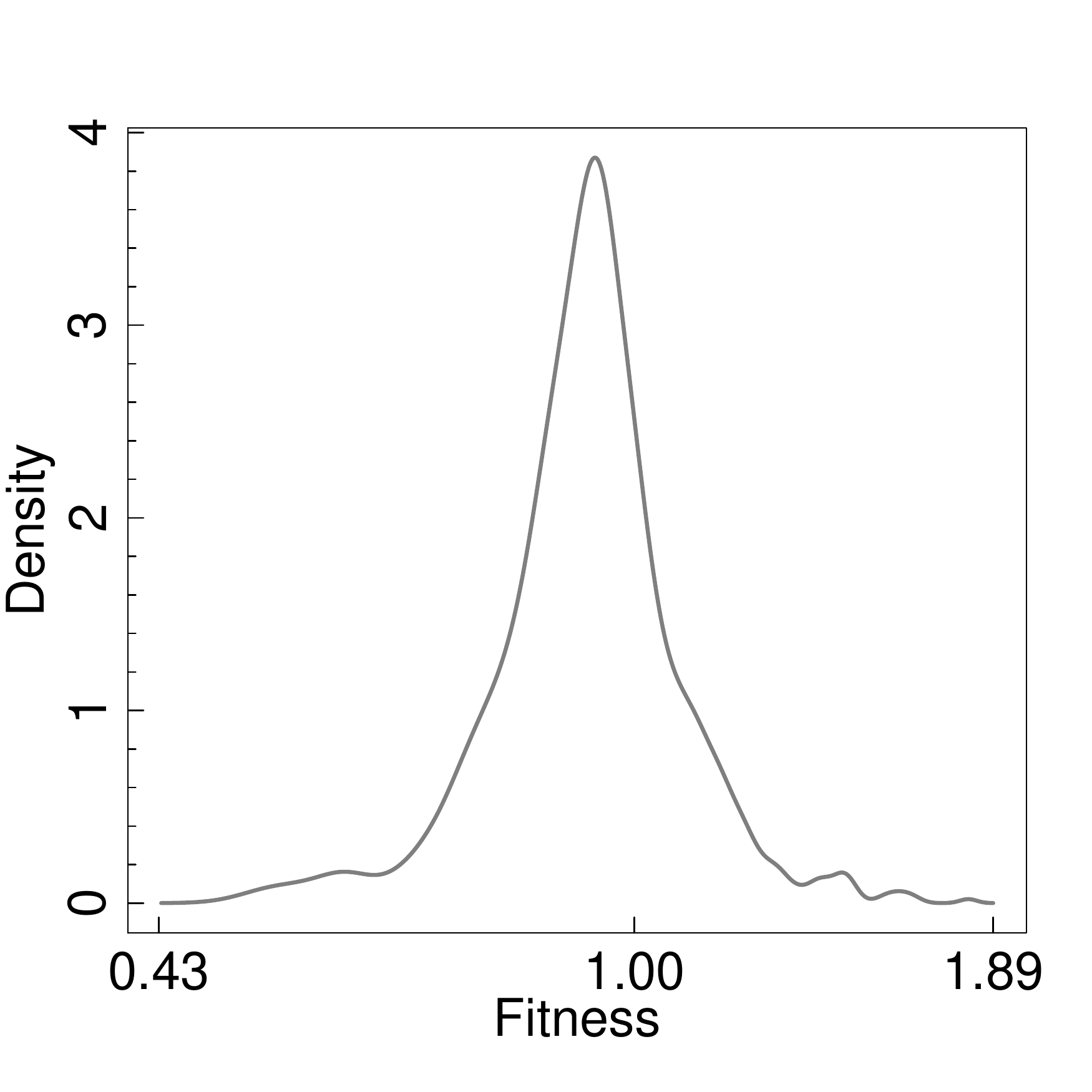}
\caption{Distribution of estimated node fitnesses.} \label{fig: pafit_fit_distribution}
\end{subfigure}
\begin{subfigure}{0.32\textwidth}
\centering
\includegraphics[width=\textwidth]{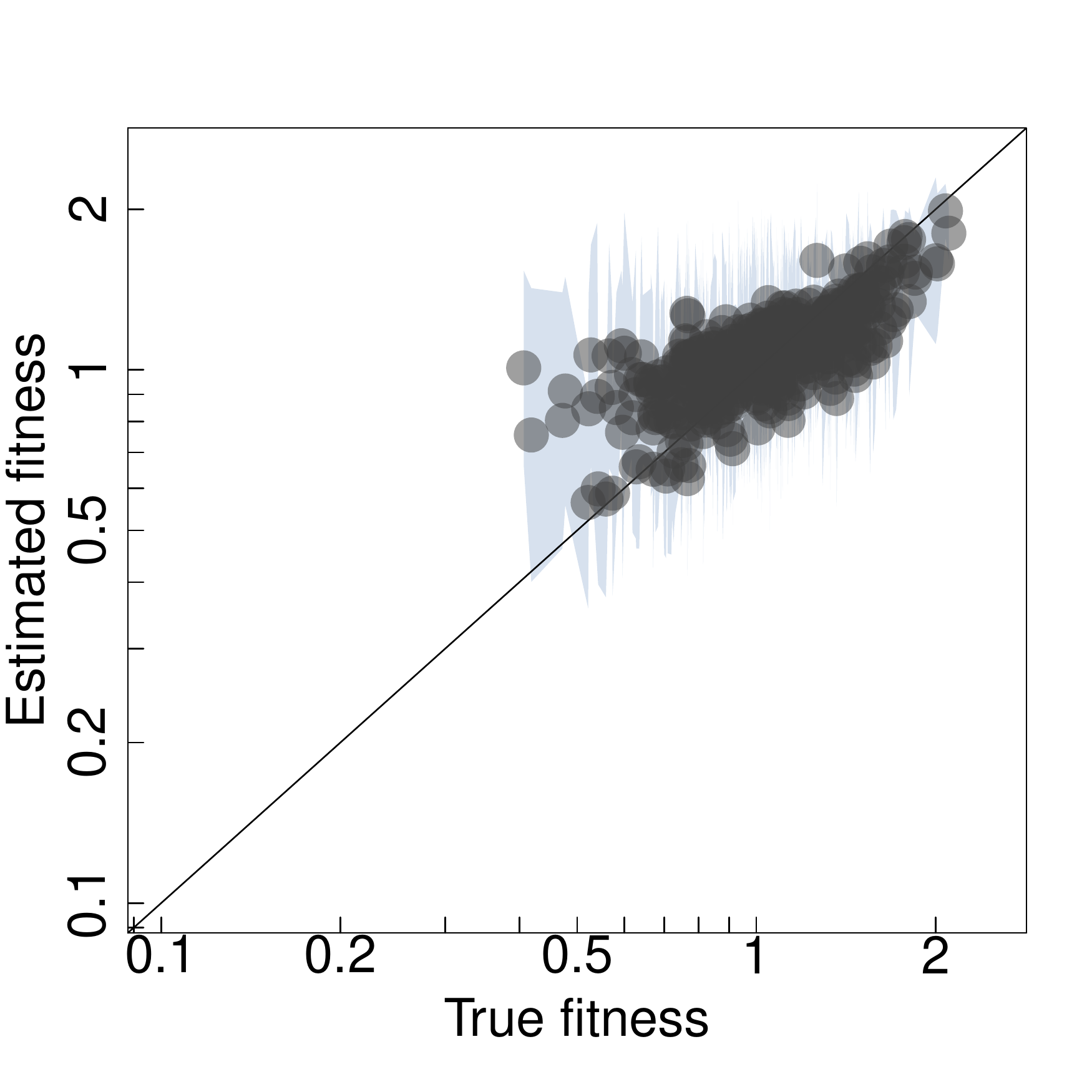} 
\caption{Estimated and true $\eta_i$ values.\\ $ $}  \label{fig: pafit_fit}
\end{subfigure}
\caption{Joint estimation of the attachment function and node fitnesses from \code{sim\_net\_3}. The red line in panel a is the true attachment function $A_k = k^{0.5}$. The true node fitnesses are sampled from a Gamma distribution with mean $1$ and inverse variance $s = 10$.\label{fig: pafit}}
\end{figure}

This time we generate a network in which the attachment function is $A_k = k^\alpha$ with $\alpha = 0.5$ and the Gamma distribution of node fitnesses has mean $1$ and variance $1/s$ with $s = 10$: 
\begin{CodeInput}
R> sim_net_3 <- generate_net(N = 1000, num_seed = 100, multiple_node = 100,
+    m = 15, s = 10, alpha = 0.5)
\end{CodeInput}                               
We then apply \code{joint_estimation}:
\begin{CodeChunk} 
\begin{CodeInput}                              
R> stats_3 <- get_statistics(sim_net_3)
R> result_PAFit <- joint_estimate(sim_net_3, stats_3)
R> summary(result_PAFit)
\end{CodeInput}
%$
\begin{CodeOutput}
Estimation results by the PAFit method. 
Mode: Both the attachment function and node fitness were estimated. 
Selected r parameter: 10 
Selected s parameter: 18.75 
Estimated attachment exponent: 0.5168941 
Attachment exponent ± 2 s.d.: (0.5097277,0.5240605 )
-------------------------------------------
Additional information: 
Number of bins: 50 
Number of iterations: 596 
Stopping condition: 0.00000001
\end{CodeOutput}
\end{CodeChunk} 
We can plot the estimated attachment function as in Figure~\ref{fig: pafit_pa}, and the distribution of the $\hat{\eta_i}$'s as in Figure~\ref{fig: pafit_fit_distribution} with the following code:

\begin{Sinput}
R> plot(result_PAFit, stats_3, min_A = 1, max_A = 40,
+    cex = 3, cex.axis = 2, cex.lab = 2)
R> lines(stats_3$center_k, stats_3$center_k^0.5, col = "red")
R> plot(result_PAFit, stats_3, plot = "f",
+    cex = 3, cex.axis = 2, cex.lab = 2)
\end{Sinput}

Concerning the estimated values, while $s$ is slightly over-estimated by $\hat{s} = 18.75$, $\hat{\alpha} = 0.52 \pm 0.01$ is a good estimate of $\alpha$. We can also plot the estimated fitnesses versus the true fitnesses as in Figure~\ref{fig: pafit_fit} with the following code:
\begin{Sinput}
R> plot(result_PAFit, stats_3, true_f = sim_net_3$fitness, plot = "true_f",
+    cex = 3, cex.axis = 2, cex.lab = 2)
\end{Sinput}
Since the mean of $\hat{\eta_i}$'s is normalized to $1$, the over-estimation of $s$ leads to over-estimation of low-value fitnesses and under-estimation of high-value fitness, as can be seen in the plot of~Figure~\ref{fig: pafit_fit}.

\begin{figure}[!h]
\centering
\begin{subfigure}{0.32\textwidth}
\centering
\includegraphics[width=\textwidth]{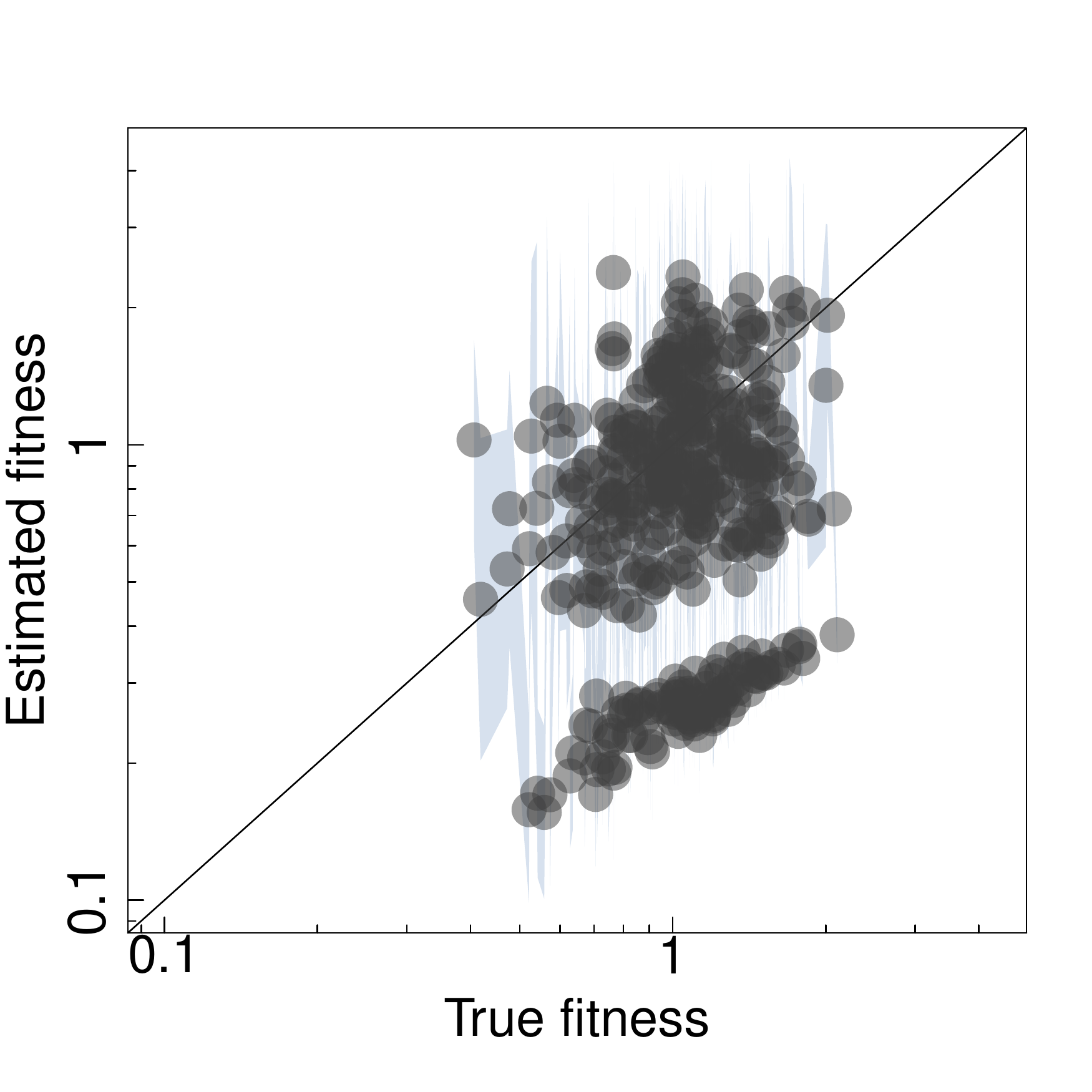} 
\caption{Estimated and true $\eta_i$ values when assuming $A_k = k$. $\hat{s}  = 3.2$.}  \label{fig: Section_5-3_fit_only_linear_pA}
\end{subfigure}
\begin{subfigure}{0.32\textwidth}
\centering
\includegraphics[width=\textwidth]{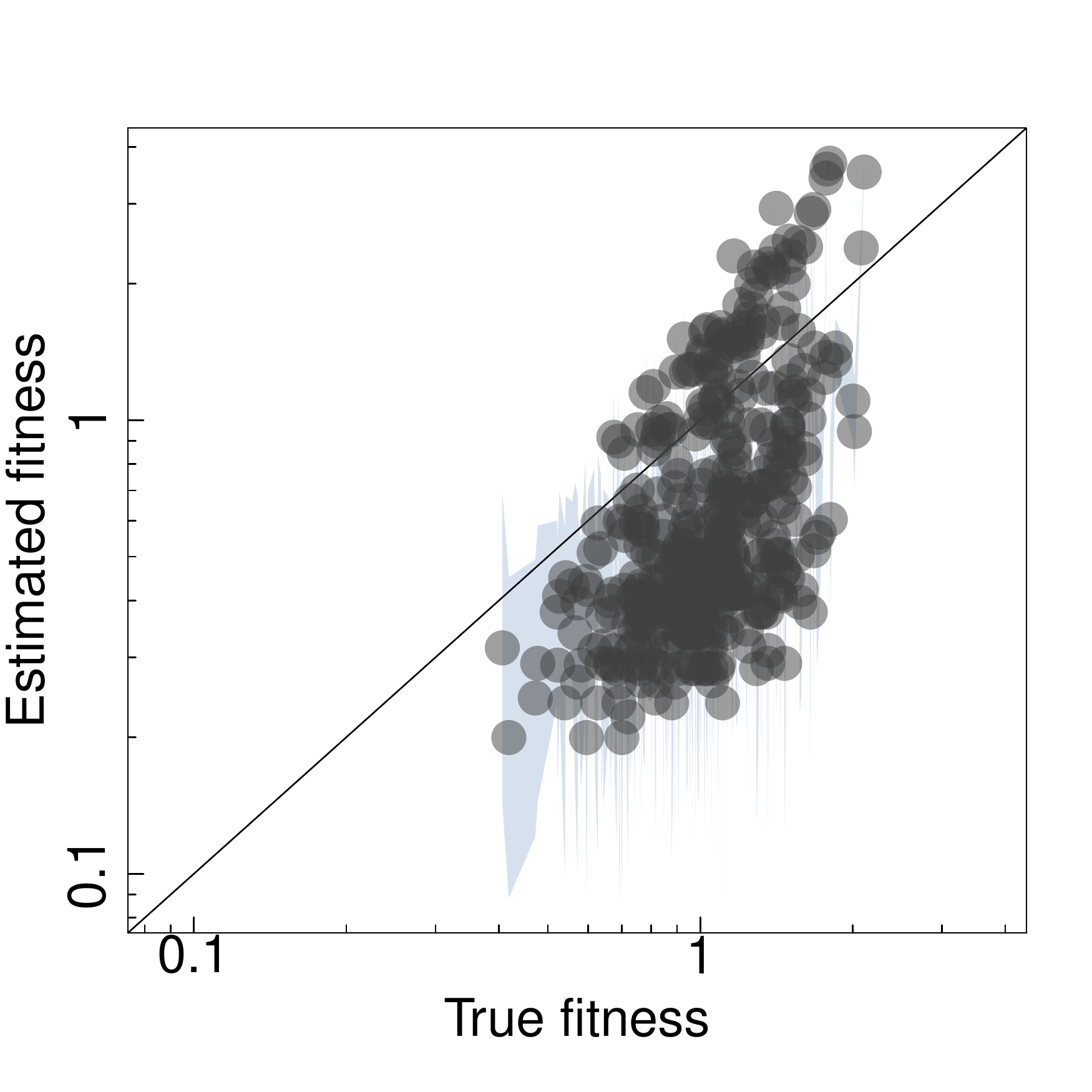}
\caption{Estimated and true $\eta_i$ values when assuming $A_k = 1$. $\hat{s}  = 4$.} \label{fig: Section_5-3_fit_only_constant_pA}
\end{subfigure}
\begin{subfigure}{0.32\textwidth}
\centering
\includegraphics[width=\textwidth]{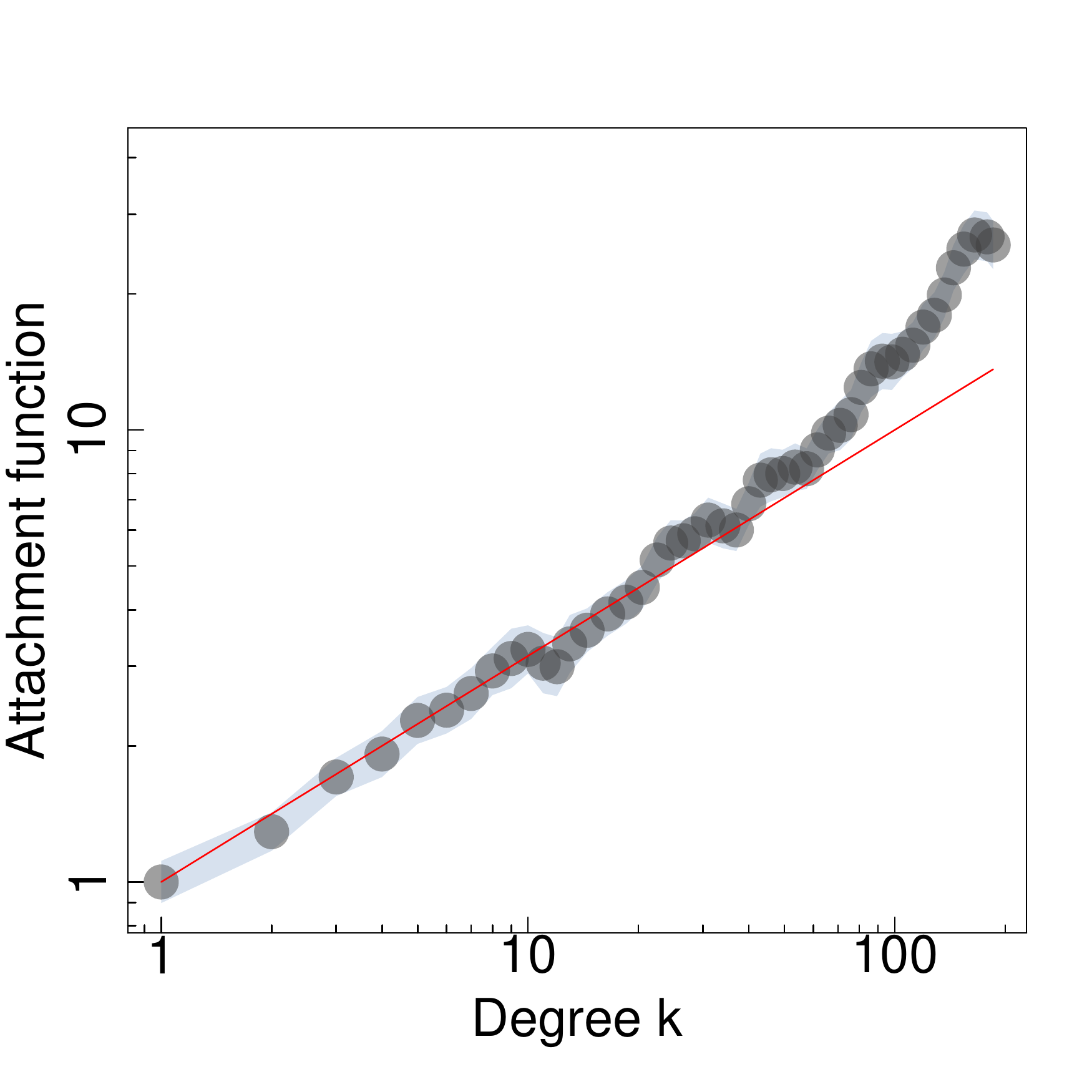}
\caption{Estimated PA function in isolation ($\hat{\alpha} = 0.60 \pm 0.05$).} \label{fig: Section_5-3_only_PA}
\end{subfigure}
\caption{Estimation of node fitnesses and the PA function in isolation from \code{sim\_net\_3}. The red line in panel c is the true attachment function $A_k = k^{0.5}$. The true node fitnesses are sampled from a Gamma distribution with mean $1$ and inverse variance $s = 10$.}  \label{fig: Section_5-3_other_model}
\end{figure}

We show how joint estimation improves on estimating either node fitnesses in isolation (Figures~\ref{fig: Section_5-3_fit_only_linear_pA} and~\ref{fig: Section_5-3_fit_only_constant_pA}) or the PA function in isolation (Figure~\ref{fig: Section_5-3_only_PA}). For estimating node fitnesses in isolation, two cases are shown: the result when we assume the BB model in which $A_k = k$ (Figure~\ref{fig: Section_5-3_fit_only_linear_pA}) and the result when we assume the Caldarelli model in which $A_k = 1$ (Figure~\ref{fig: Section_5-3_fit_only_constant_pA}). In either case, the estimated node fitnesses are visually worse than those of the joint estimation in Figure~\ref{fig: pafit_fit}. Similarly, estimating the PA function in isolation apparently led to overestimation of the PA function in the region of large $k$. To conclude, estimating either node fitnesses or the PA function in isolation would likely be worse than the joint estimation, if the underlying assumptions about the true node fitnesses and the true PA function are~wrong.

\section{Simulation Study}
\label{sec: simulation_study}
In this section, we present the results of a simulation study that we conducted to assess the performance of the \code{joint_estimation} function. We assume the functional form $A_k = k^\alpha$ for the attachment function. To cover the spectrum of PA and anti-PA phenomena, we choose four values for $\alpha$:  $-0.5$, $0$, $0.5$, and $1$. We sample node fitnesses from a Gamma distribution with mean $1$ and variance $1/s$. Three values for~$s$ we chose are: $5$, $20$, and $80$. While small values of $s$ lead to widely varied node fitnesses, large values of $s$ leads to highly concentrated node fitnesses.  

  For each combination of $\alpha$ and $s$, we generated $M = 50$ networks, and estimated $A_k$ and $s$ for each network using the \code{joint_estimation} function. We then fit the form $\hat{A}_k = k^\alpha$ to $\hat{A_k}$ in order to estimate $\alpha$. We then compared the means of the estimation results of $\alpha$ and $s$ with the true values. Each simulated network has a total of $1000$ nodes, where the initial graph has $200$ nodes and $50$ new nodes are added at each time-step for a total of $10$ time-steps. Each new node has $50$ new edges.

\begin{figure}[!h]
\centering
\includegraphics[width=\textwidth]{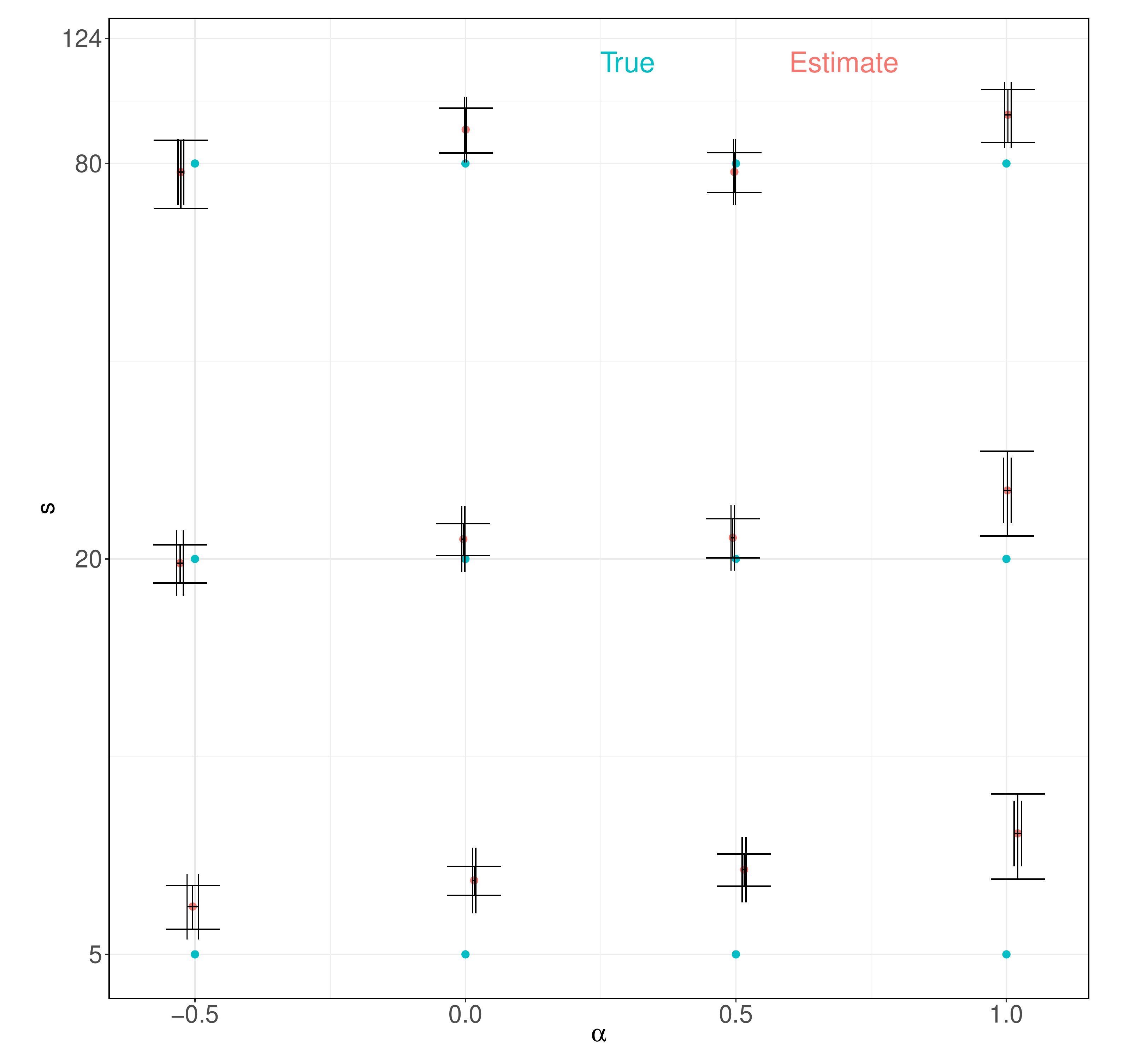}
\caption{Simulation result. For each combination of $\alpha$ and $s$, we generated $50$ networks using attachment function $A_k = k^\alpha$ and a Gamma distribution with mean $1$ and inverse variance~$s$ for node fitnesses. Each red point indicates the average of the esimated $\alpha$ and the selected $s$ over $100$ simulations for the corresponding combination of $\alpha$ and $s$. At each red point, the horizontal/verticalcal bar indicates the interval of the estimated $\alpha$/selected $s$ $\pm$ 2 s.d., respectively. \label{fig: estimation_result}}
\end{figure}

The results are shown in Figure~\ref{fig: estimation_result}. The attachment exponent $\alpha$ was estimated reasonably well across all combinations of $\alpha$ and $s$. Except for the cases in which the attachment function grows fast ($\alpha = 1$) or the case in which node fitnesses have high variance ($s = 5$), the estimated values of $s$ were also acceptable. The case of $s = 5$ resulted in a slight over-estimation, which is perhaps attributable to high node fitnesses variance. We also notice that $s$ was slightly over-estimated when $\alpha = 1$, which may be caused by the fast growing rate of the PA function. One also notices that the intervals of $\hat{s}$ $\pm$ 2 s.d. are much larger than those for $\hat{\alpha}$. The above observations imply that it is much harder to estimate $s$ than $\alpha$.

\section{Analysis of a collaboration network between scientists}
\label{sec: real_example}
 In this section, we demonstrate the complete work-flow for the joint estimation of $A_k$ and $\eta_i$ on a collaboration network between scientists from the field of complex networks. In this network, nodes represent scientists and an undirected edge exists between them if, and only if, they have coauthored a paper. The degree of a node represents the number of collaborators of a scientist, since multiple edges are not considered. The temporal network is stored in \code{coauthor.net}, and the names of the scientists are stored in \code{coauthor.author_id}. The network without timestamps was compiled by Mark Newman from the bibliographies of two review articles on complex networks~\citep{newman_community}. Paul Sheridan, the second author of the present work, augmented the dataset with time stamps. More information on the dataset can be found in the package reference manual. 
 
\begin{figure}[!h]
\centering 
\includegraphics[width=0.75\textwidth]{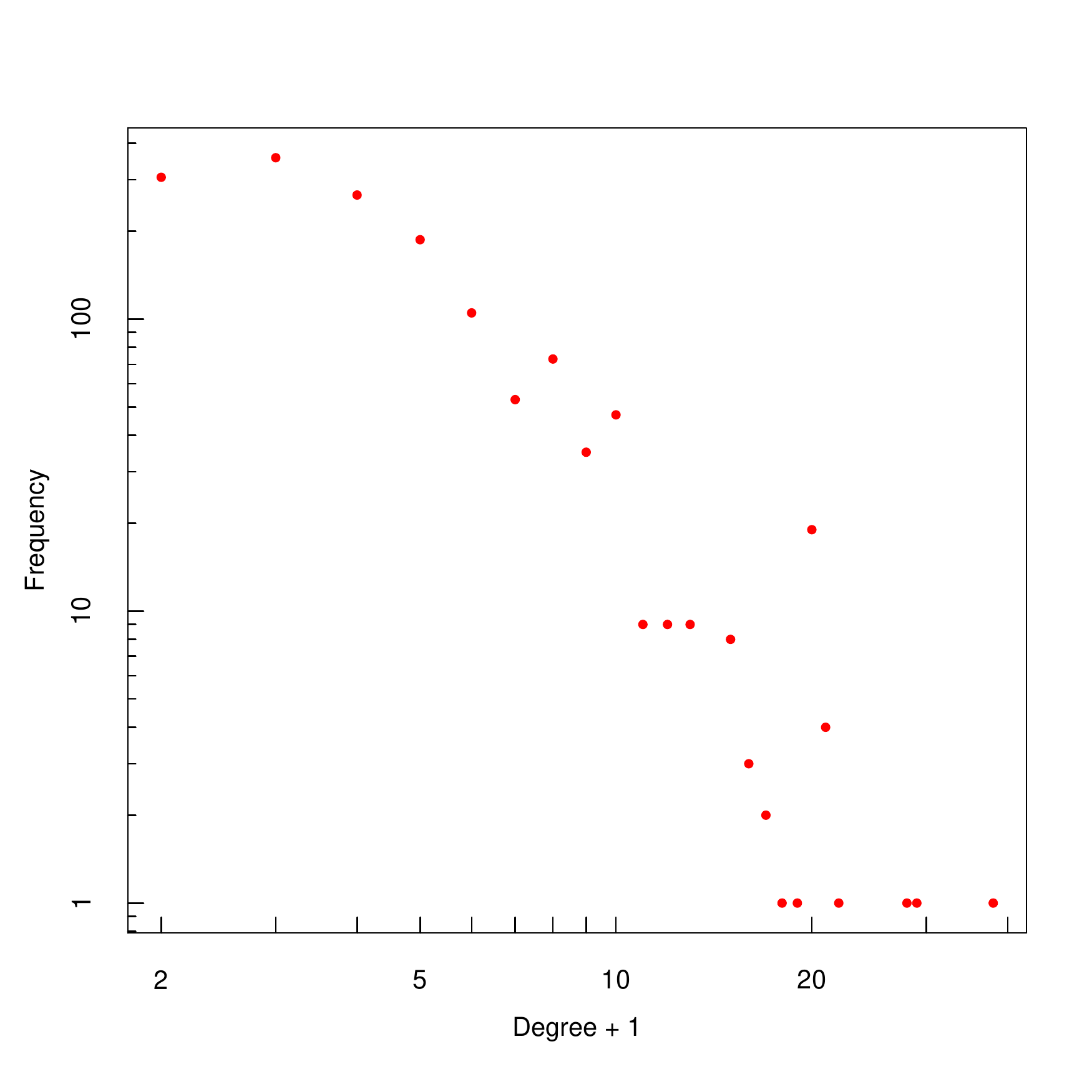} 
\caption{Degree distribution of the final snapshot of the collaboration network.} \label{fig: collaboration_degree_dist}
\end{figure}

The first step in the analysis is to convert the edge-list matrix \code{coauthor.net} to a \code{PAFit\_net} object, and get the summary statistics using the function \code{get\_statistics}.
\begin{CodeChunk} 
\begin{CodeInput}
R> set.seed(1)
R> true_net <- as.PAFit_net(coauthor.net, type = "undirected")
R> net_stats <- get_statistics(true_net)
R> summary(net_stats)
\end{CodeInput}
%$
\begin{CodeOutput}
Contains summary statistics for the temporal network. 
Type of network: undirected 
Number of nodes in the final network: 1498 
Number of edges in the final network: 5698 
Number of new nodes: 1358 
Number of new edges: 1255 
Number of time-steps: 145 
Maximum degree: 37 
Number of bins: 38 
\end{CodeOutput}
\end{CodeChunk}
The temporal network grew in $145$ time-steps from an initial network at September 2000, to a final state at September 2007. The resolution of those time-steps is monthly. The final network has $1498$ scientists with $5698$ collaborations among them. One can plot the degree distribution of the final snapshot as follows:
\begin{Sinput}
R> plot(true_net,plot = "degree")
\end{Sinput}
This will produce the plot of Figure~\ref{fig: collaboration_degree_dist}.

Before any estimation of the PA function and node fitnesses is carried out, one can test whether the linear PA-only case is consistent with the observed degree distribution of the collaboration network:
\begin{Sinput}
R> test_linear_PA_result <- test_linear_PA(net_stats$final_deg)
R> print(test_linear_PA_result)
\end{Sinput}
This will generate Table~\ref{tab: test_linear_PA_result}. In this case, since the Yule and Waring distributions are not the best models, one can conclude that the linear PA-only case is inconsistent with the observed degree vector.
\begin{table}[!h]
\centering
\begin{tabular}{ l l l l}
\hline
Model & Log-likelihood & AIC  & BIC\\ \hline
\code{nb}     &-2415.08 & 4840.16 & 4866.72\\ 
\code{pois}   &-2468.63 &4945.27 & 4966.51 \\ 
\code{waring} &-2557.49       &5124.99 &5151.55 \\ 
\code{geom}   &-2898.64       &5811.27 &5848.45 \\ 
\code{yule} &-2959.97      &5929.95 &5956.51 \\ 
\hline
\end{tabular}
\caption{The result of applying the function \code{test\_linear\_PA} to the observed degree vector of the collaboration network. This function calculates the AIC and BIC of five models: Yule (\code{yule}), Waring (\code{waring}), Poisson (\code{pois}), geometric (\code{geom}), and negative binomial (\code{nb}) when fitting them to the observed degree vector. {\label{tab: test_linear_PA_result}}}
\end{table}

\begin{figure}[!h]
\centering
\begin{subfigure}{0.49\textwidth}
\centering
\includegraphics[width=\textwidth]{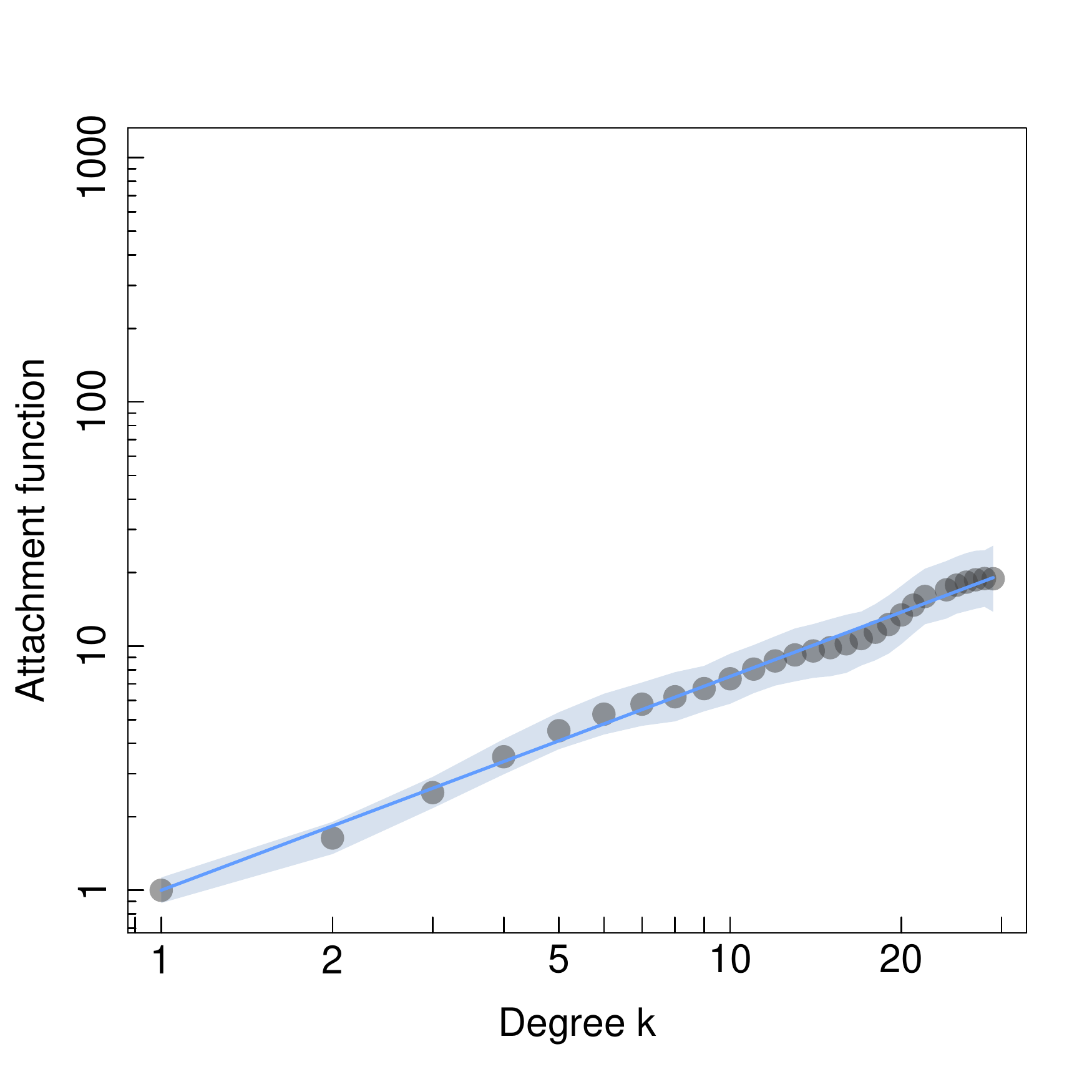} 
\caption{Estimated $A_k$ with joint estimation by PAFit ($\hat{\alpha} = 1.00 \pm 0.05$).}  \label{fig: complex_PAFit_PA}
\end{subfigure}
\begin{subfigure}{0.49\textwidth}
\centering
\includegraphics[width=\textwidth]{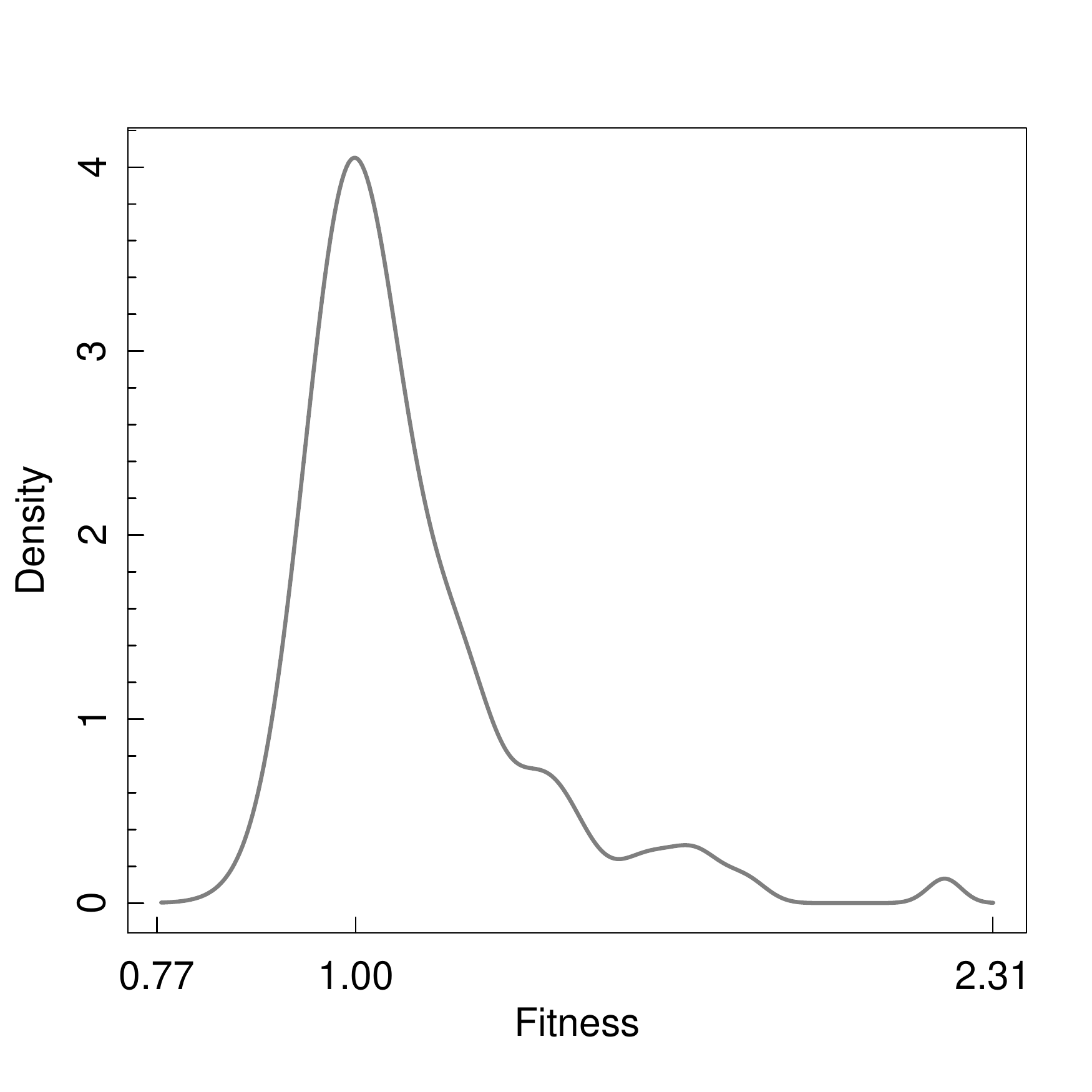}
\caption{Histogram of estimated node fitnesses. \\ $ $} \label{fig: complex_histogram}
\end{subfigure}
\begin{subfigure}{0.32\textwidth}
\centering
\includegraphics[width=\textwidth]{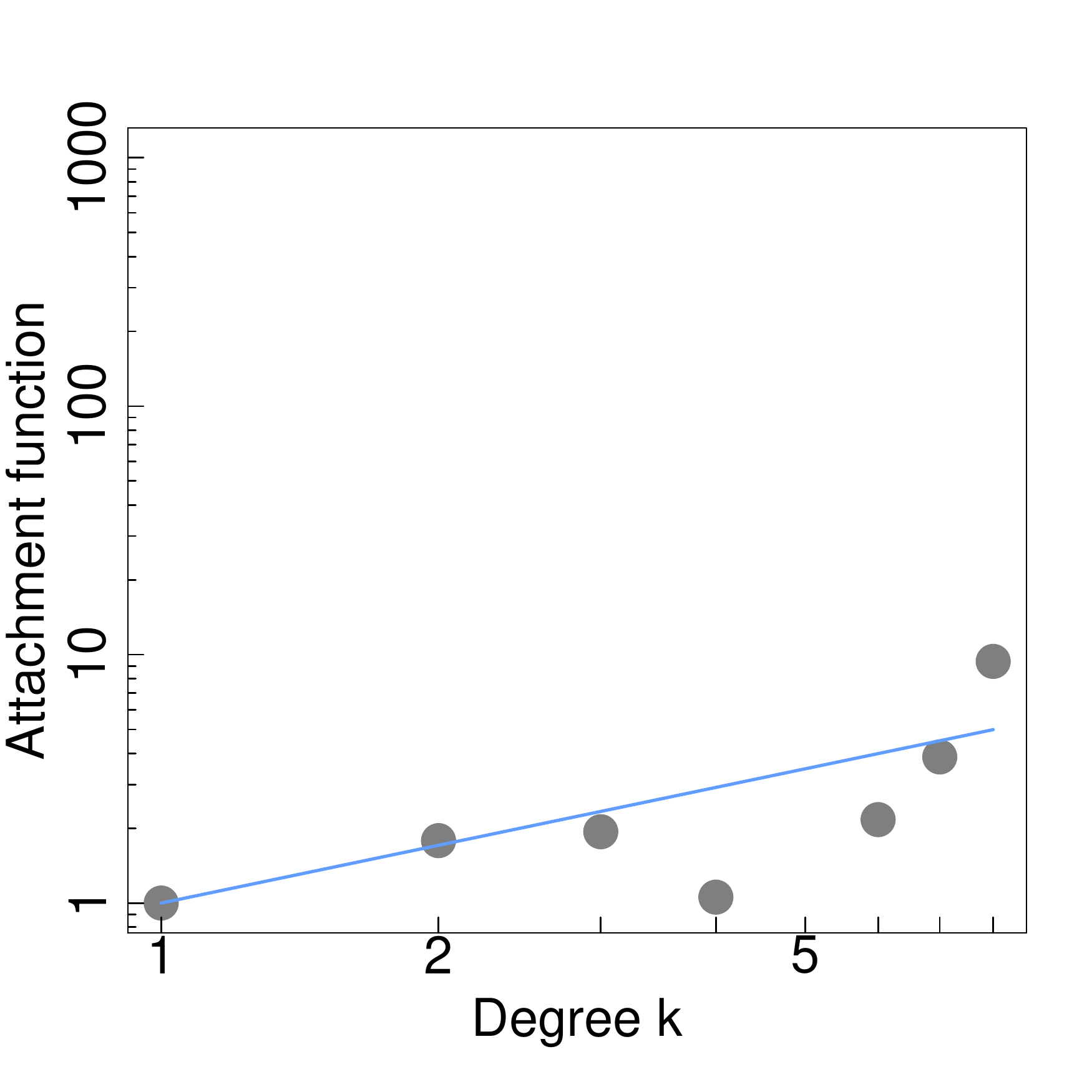}
\caption{Jeong's method \\ ($\hat{\alpha} = 0.77 \pm 0.80$).} \label{fig: complex_Jeong}
\end{subfigure}
\begin{subfigure}{0.32\textwidth}
\centering
\includegraphics[width=\textwidth]{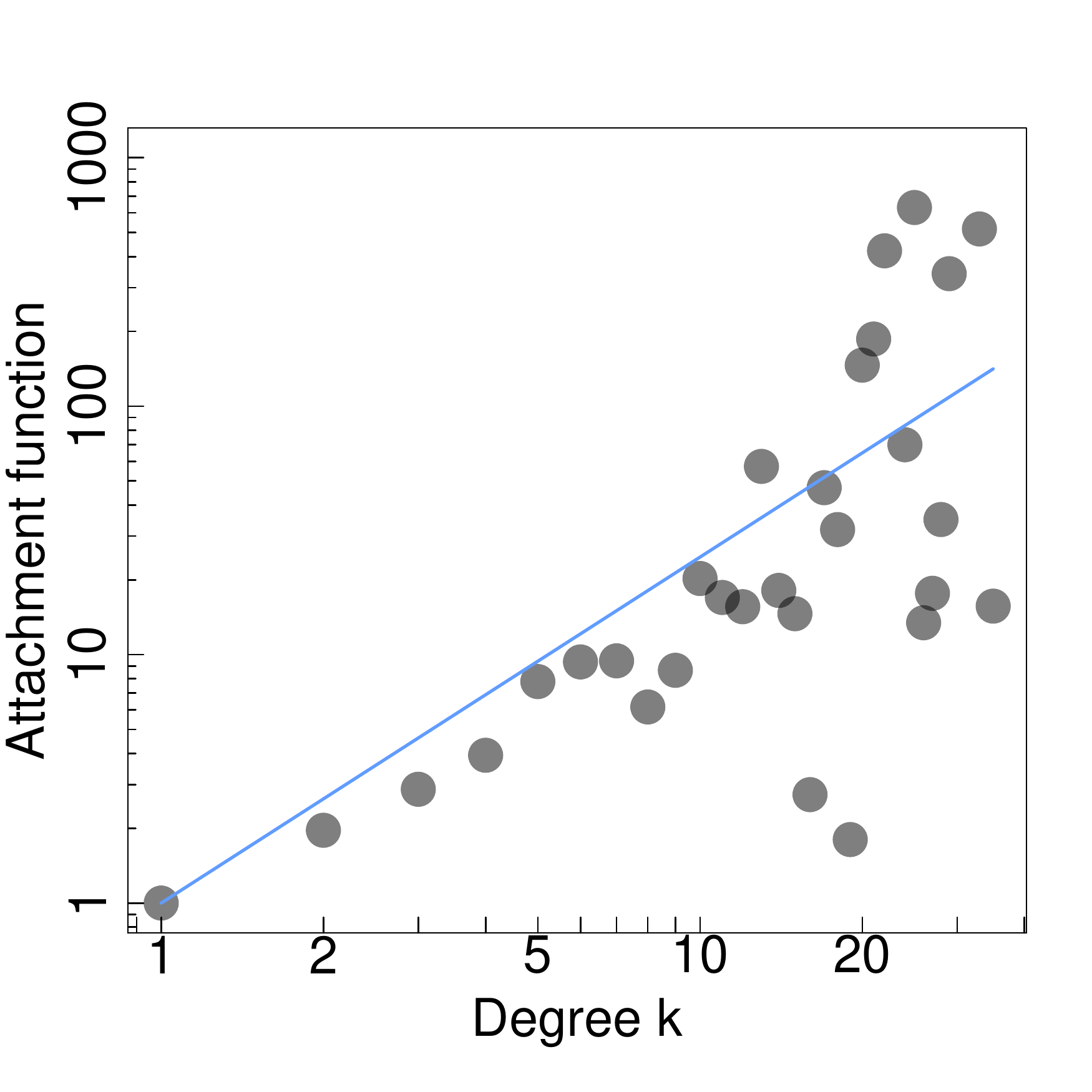} 
\caption{Newman's method \\ ($\hat{\alpha} = 1.39 \pm 0.56$).}  \label{fig: complex_Newman}
\end{subfigure}
\begin{subfigure}{0.32\textwidth}
\centering
\includegraphics[width=\textwidth]{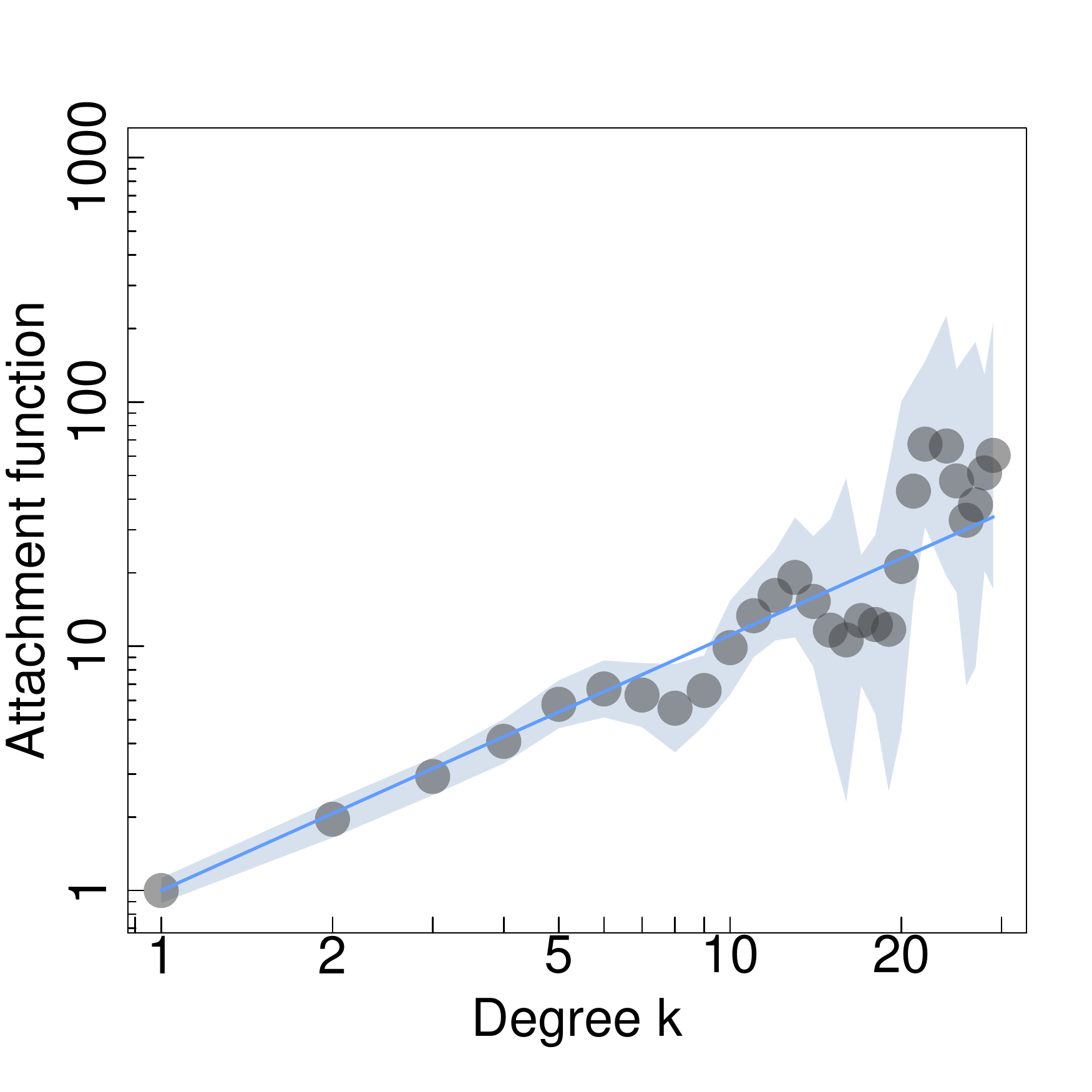}
\caption{PAFit \\ ($\hat{\alpha} = 1.05 \pm 0.07$).} \label{fig: complex_onlyA}
\end{subfigure}
\caption{Estimation of the attachment function and node fitnesses for the network scientist collaboration network. Panels a and b show the joint estimation result, while panels c, d, and e show the results when we estimated the PA function in isolation.}  \label{fig: complex_all_cases}
\end{figure}

To further investigate the PA function and node fitnesses, we invoke the \code{joint\_estimate} function for joint estimation:
\begin{CodeChunk} 
\begin{CodeInput}
R> full_result <- joint_estimate(true_net, net_stats)
R> summary(full_result)
\end{CodeInput}
%$
\begin{CodeOutput}
Estimation results by the PAFit method. 
Mode: Both the attachment function and node fitness were estimated. 
Selected r parameter: 10 
Selected s parameter: 45 
Estimated attachment exponent: 0.9951764 
Attachment exponent ± 2 s.d.: (0.9715202,1.018833)
-------------------------------------------
Additional information: 
Number of bins: 38 
Number of iterations: 607 
Stopping condition: 0.00000001
\end{CodeOutput}
\end{CodeChunk}
We can visualize the estimated attachment function and the distribution of estimated node fitnesses by:
\begin{Sinput}
R> plot(full_result, net_stats, line = "TRUE",
+    cex = 2, cex.axis = 1.5, min_A = 1, max_A = 1000, cex.lab = 1.5)
R> plot(full_result, net_stats, plot = "f",
+    cex = 2, cex.axis = 1.5, cex.lab = 1.5)
\end{Sinput}
This snippet will sequentially generate the plots of Figures~\ref{fig: complex_PAFit_PA} and~\ref{fig: complex_histogram}. When other options are set at their default values, the option \code{line = "TRUE"} will plot the function $\hat{A}_k = k^{\hat{\alpha}}$, which is a straight line on a double logarithmic scale.

The best fit model when we performed joint estimation is close to the BB model. In Figure~\ref{fig: complex_PAFit_PA}, the estimated $A_k$ is an increasing function with $\hat{\alpha} = 1.00 \pm 0.05$. We take this as evidence in favor of the presence of linear PA in the collaboration network. Let us take a concrete example: a network scientist with twenty collaborators has roughly twice the chance to get a new collaborator compared with someone who only has ten collaborators, assuming they have the same fitness. For comparison's sake, we also plot the estimation results of $A_k$ in isolation using Jeong's method, Newman's method, and PAFit in Figures~\ref{fig: complex_Jeong},~\ref{fig: complex_Newman}, and~\ref{fig: complex_onlyA}, respectively:
\begin{Sinput}
R> result_Jeong <- Jeong(true_net, net_stats)
R> result_Newman <- Newman(true_net, net_stats)
R> result_onlyA <- only_A_estimate(true_net, net_stats)
R> plot(result_Jeong, net_stats, line = "TRUE", min_A = 1, max_A = 1000,
+    cex = 3, cex.axis = 2, cex.lab = 2)
R> plot(result_Newman, net_stats, line = "TRUE", min_A = 1, max_A = 1000,
+    cex = 3, cex.axis = 2, cex.lab = 2)
R> plot(result_onlyA, net_stats, line = "TRUE", min_A = 1, max_A = 1000,
+    cex = 3, cex.axis = 2, cex.lab = 2)
\end{Sinput}
The options \code{min\_A = 1} and \code{max\_A = 1000} specify the range of the vertical axis and are needed for making the plots comparable.

The high variance of $\hat{\alpha}$ from either Jeong's method or Newman's method would render qualitative assessments of the PA function inconclusive, if one relied only on those methods: one could not confidently ascertain whether the PA function is sub-linear, linear, or super-linear in nature. We notice that the estimated $A_k$ obtained from the joint estimation resembles that of Figure~\ref{fig: complex_onlyA}, when we estimate it in isolation. The reason is that estimated node fitnesses in Figure~\ref{fig: complex_histogram} are highly concentrated around the mean. Thus their distribution is not very far from the case when all the fitnesses are $1$. Nevertheless, we observe that the estimated $A_k$ from the joint estimation is reduced when compared with that of Figure~\ref{fig: complex_onlyA}. This is expected since in the joint estimation, a portion of the connection probability in Equation~\ref{eq: model} is explained by node fitness.

Although the distribution in the plot of Figure~\ref{fig: complex_histogram} is concentrated around its mean, we notice that its right tail is rather long, which is a sign that this tail contains interesting information. We can extract the information from this region by finding the topmost `fittest' network scientists. This can be done as follows:
\begin{Sinput} 
R> sorted_fit <- sort(full_result$estimate_result$f, decreasing = TRUE)
R> top_scientist <- coauthor.author_id[names(sorted_fit), ]
R> print(cbind(sorted_fit[1:10], top_scientist[1:10, 2]))
\end{Sinput}
This snippet will produce the results show in Table~\ref{tab: top_ten}. The table shows the ten network scientists that we found to have the highest ability to attract new collaborators from the field.
\begin{table}[!h]
\centering
\begin{tabular}{ l l l}
\hline
Rank & Estimated fitness & Name \\ \hline
 1 & 1.42 & BARABASI, A \\ 
2  & 1.35 &  NEWMAN, M \\ 
3  & 1.26 & JEONG, H     \\ 
4 & 1.25 & LATORA, V     \\ 
5  & 1.24 & ALON, U      \\  
6 & 1.23 &   OLTVAI, Z  \\ 
7 & 1.23 & YOUNG, M \\
8 & 1.22 &  WANG, B \\
9  & 1.21 & SOLE, R         \\ 
10  & 1.21 & BOCCALETTI, S \\
\hline
\end{tabular}
\caption{Ten topmost `fittest' network scientists in the field of complex networks. {\label{tab: top_ten}}}
\end{table}
Anyone acquainted with the field will recognized a number of familiar faces. For example, at the top of the list is none other than Albert-L{\'a}szl{\'o} Barab{\'a}si, who introduced the BA model. Number two and number three are Mark Newman and Hawoong Jeong, who respectively are the authors of the eponymously named Newman's method and Jeong's method.

\section{Conclusion}
\label{sec: final}
We introduced the \proglang{R} package \pkg{PAFit}, which provides a comprehensive framework for the non-parametric estimation of PA and node fitness mechanisms in the growth of temporal complex networks. In summary, \pkg{PAFit} implements functions to simulate various temporal network models based on these two mechanisms, gathers summary statistics from real-world temporal network datasets, and estimates non-parametrically the attachment function and node fitnesses. We provided a number of simulated examples, as well as a complete analysis of a real-world collaboration network.

\section*{Acknowledgments}
This work was supported in part by grants from the Japan Society for the Promotion of Science KAKENHI [JP16J03918 to T.P. and 16H01547 to H.S.].

\bibliography{jss3094.bib}

\end{document}